\documentclass[12pt]{elsarticle}

\usepackage[margin=1.1in]{geometry}

\usepackage{amsmath, amssymb}
\usepackage{bm}
\usepackage{mathrsfs}
\usepackage{graphicx, fancyhdr}
\usepackage{placeins} 
\usepackage{floatrow}

\usepackage{ifpdf}
\usepackage{multirow}
\usepackage{caption}
\usepackage{subcaption}

\newdimen\figrasterwd
\figrasterwd\textwidth

\usepackage{setspace}
\usepackage{xcolor}

\usepackage{hyperref}
\hypersetup{
 colorlinks=true,
 linkcolor=blue,
 citecolor=blue,
 urlcolor=blue
 }

\begin{document}

\begin{frontmatter}

\title{Graph Network Surrogate Model for Optimizing the Placement of Horizontal Injection Wells for CO$_2$ Storage}

\author{Haoyu Tang\corref{mycorrespondingauthor}}
\ead{hytang@stanford.edu}

\author{Louis J.~Durlofsky}
\ead{lou@stanford.edu}
\address{Department of Energy Science \& Engineering, Stanford University, Stanford, CA 94305, USA}

\cortext[mycorrespondingauthor]{Corresponding author}

\begin{abstract}
Optimizing the locations of multiple CO$_2$ injection wells will be essential as we proceed from demonstration-scale to large-scale carbon storage operations. Well placement optimization is, however, a computationally intensive task because the flow responses associated with many potential configurations must be evaluated. There is thus a need for efficient surrogate models for this application. In this work we develop and apply a graph network surrogate model (GNSM) to predict the global pressure and CO$_2$ saturation fields in 3D geological models for arbitrary configurations of four horizontal wells. The GNSM uses an encoding-processing-decoding framework where the problem is represented in terms of computational graphs. Separate networks are applied for pressure and saturation predictions, and a multilayer perceptron is used to provide bottom-hole pressure (BHP) for each well at each time step. The GNSM is shown to achieve median relative errors of 4\% for pressure and 6\% for saturation over a test set involving very different plume shapes and dynamics. Speedup is about a factor of $120\times$ relative to high-fidelity simulation. The GNSM is applied for optimization using a differential evolution algorithm, where the goal is to minimize the CO$_2$ footprint subject to constraints on the well configuration, plume location and well BHPs. Optimization results using the GNSM are shown to be comparable to those achieved using (much more expensive) high-fidelity simulation. 

\end{abstract}

\end{frontmatter}


\section{Introduction}
\label{sec:intro}

Computational optimization will be a key technology for the design and expansion of CO$_2$ storage operations. These optimizations could entail the determination of well locations, or the joint determination of well locations and control settings (e.g., rates or pressures), such that a particular metric is minimized. Example metrics to minimize include the size of the CO$_2$ footprint, the fraction of CO$_2$ that remains mobile, or monitoring cost. Optimization can be computationally expensive, however, due to the need to evaluate the large number of configurations proposed by the optimization algorithm. In a conventional optimization procedure, each of these evaluations involves a full flow simulation. In surrogate modeling procedures, the function evaluations required during optimization are performed using an approximate but fast proxy model. Our goal in this work is to develop a deep-learning-based graph network surrogate model that is applicable for well placement optimization in CO$_2$ storage.

The optimization of well placement and control for geological CO$_2$ storage has been addressed in many studies. Cameron and Durlofsky~\citep{cameron2012optimization} utilized a direct search algorithm to minimize mobile CO$_2$ for injection scenarios involving a set of horizontal wells. Cihan et al.~\citep{cihan2015optimal} employed constrained differential evolution (CDE) to optimize well locations and injection rates. Other optimization algorithms, including iterative Latin hypercube sampling (ILHS)~\citep{goda2013global}, genetic algorithms (GAs)~\citep{stopa2016optimization}, particle swarm optimization (PSO), and differential evolution (DE)~\citep{zou2023integrated}, have also been applied for storage problems. The potential for induced seismicity effects has been considered in storage optimization. Zheng et al.~\citep{zheng2021geologic} applied a controlled nondominated sorting GA (NSGA-II) algorithm for multiobjective optimization with coupled flow and geomechanics. In their setup, vertical well locations were determined to maximize the amount of CO$_2$ stored while minimizing geomechanical risks. In recent work, DE and PSO were shown to perform well for storage optimization problems involving multiple deviated or horizontal wells and realistic constraints~\citep{zou2023integrated}. Arouri and Sayyafzadeh~\citep{arouri2022_Comp_Geo} successfully applied adaptive moment estimation with stochastic gradient approximation for optimizing deviated wells in oil field problems. This and related optimization algorithms are also suitable for use in carbon storage. 

There has been extensive work on developing machine learning surrogates for modeling CO$_2$ storage and other subsurface flow operations. Convolutional neural networks (CNNs), particularly those involving U-Net architectures, have been applied by, e.g., Wen et al.~\citep{wen2021towards}, Tang et al.~\citep{tang2022deep}, Zhao et al.~\citep{zhao2023efficient}, Wang et al.~\citep{wang2024deep}, Molokwu et al.~\citep{molokwu2024application}, and Feng et al.~\citep{feng2024uncertainty}. Tang et al.~\citep{tang2022deep} implemented a 3D recurrent residual U-Net, which was applied to simulate (and history match) carbon storage with coupled flow and geomechanics. Han et al.~\citep{han2024surrogate} extended this approach for use with hierarchical Markov chain Monte Carlo history matching (for flow-only CO$_2$ storage problems).
Continuous conditional generative adversarial networks (CCGANs) were used by Stepien et al.~\cite{stepien2023continuous} for high-accuracy CO$_2$ plume predictions. Fourier neural operators (FNO) and deep operator networks (DeepONet) have been used in geological CO$_2$ storage applications by Yan et al.~\citep{yan2022robust}, Diab and Al-Kobaisi~\citep{diab2023u}, Witte et al.~\citep{witte2023fast}, and Seabra et al.~\citep{seabra2024ai}. 

Machine learning surrogate models, of various types, are well suited for use in optimization of CO$_2$ storage and related problems. These methods are particularly effective when many expensive function evaluations (forward simulations) must be performed, as is the case with population-based optimization methods such as PSO and DE. In the context of well control optimization for carbon storage, Chen et al.~\cite{chen2024optimization} integrated the embed to control and observe (E2CO) method with reinforcement learning to improve storage efficiency. You et al.~\cite{you2020machine} used two machine-learning methods, artificial neural networks (ANNs) and radial basis neural networks, for a CO$_2$--enhanced oil recovery (EOR) project. The goal in that work was the co-optimization of CO$_2$ storage and hydrocarbon recovery. Vaziri and Sedaee~\cite{vaziri2024application} also adopted ANNs to co-optimize CO$_2$ storage and breakthrough time. 

In the context of well placement optimization, Esfandi et al.~\cite{esfandi2024effect} explored the application of boosting-based algorithms for CO$_2$--EOR projects in a 3D light oil carbonate reservoir with vertical wells. They evaluated the performance of a number of boosting algorithms, including AdaBoost, CatBoost, Gradient Boosting, LightGBM, and XGBoost. Musayev et al.~\cite{musayev2023optimization} used an ANN-based surrogate to optimize the location of one vertical CO$_2$ injection well and one brine production well. The goal in that work was to maximize the cumulative amount of CO$_2$ stored at the end of the injection period. Fotias et al.~\cite{fotias2024optimization} introduced a Bayesian optimization framework to maximize the amount of injected CO$_2$ in a system involving multiple vertical wells. They used a Gaussian-process-based surrogate to account for model uncertainty.

In this paper we develop a deep-learning-based surrogate, specifically a graph network surrogate model (GNSM), that is applicable for use in well placement optimization in CO$_2$ storage. The GNSM procedure developed in this study represents a significant extension of an earlier method introduced for oil-water problems~\citep{tang2024graph}. In that work 2D (unstructured) models, with each well perforated in only a single grid cell, were treated. The models considered here are 3D (structured), with horizontal wells that are each perforated in multiple grid blocks. We construct separate networks to predict the full 3D pressure and saturation fields at a sequence of time steps. An additional multilayer perceptron is trained to provide BHPs for each well. The GNSM is used with a differential evolution optimization algorithm to minimize the CO$_2$ footprint resulting from 40~Mt of total injection from four horizontal wells over a 20-year period.

This paper proceeds as follows. The geological model used for our assessments is described in Section~\ref{sec:problem_setup_3d}. The detailed GNSM framework, which includes new features and treatments relative to those used for 2D cases, is presented in Section~\ref{sec:gnsm_3d}. In Section~\ref{sec:result_test_3d}, we provide GNSM predictions for a set of test cases involving various configurations of horizontal CO$_2$ injection wells. The application of the trained GNSM for well placement optimization, with the goal of minimizing CO$_2$ footprint, is presented in Section~\ref{sec:result_opt_3d}. We summarize this work and provide suggestions for future research in Section~\ref{sec:summary}.


\section{Problem setup}
\label{sec:problem_setup_3d}    
\begin{figure}[H]
	\centering
	\subfloat[Overall model including the central storage aquifer (white square) and the large surrounding region]{\includegraphics[width=0.45\textwidth]{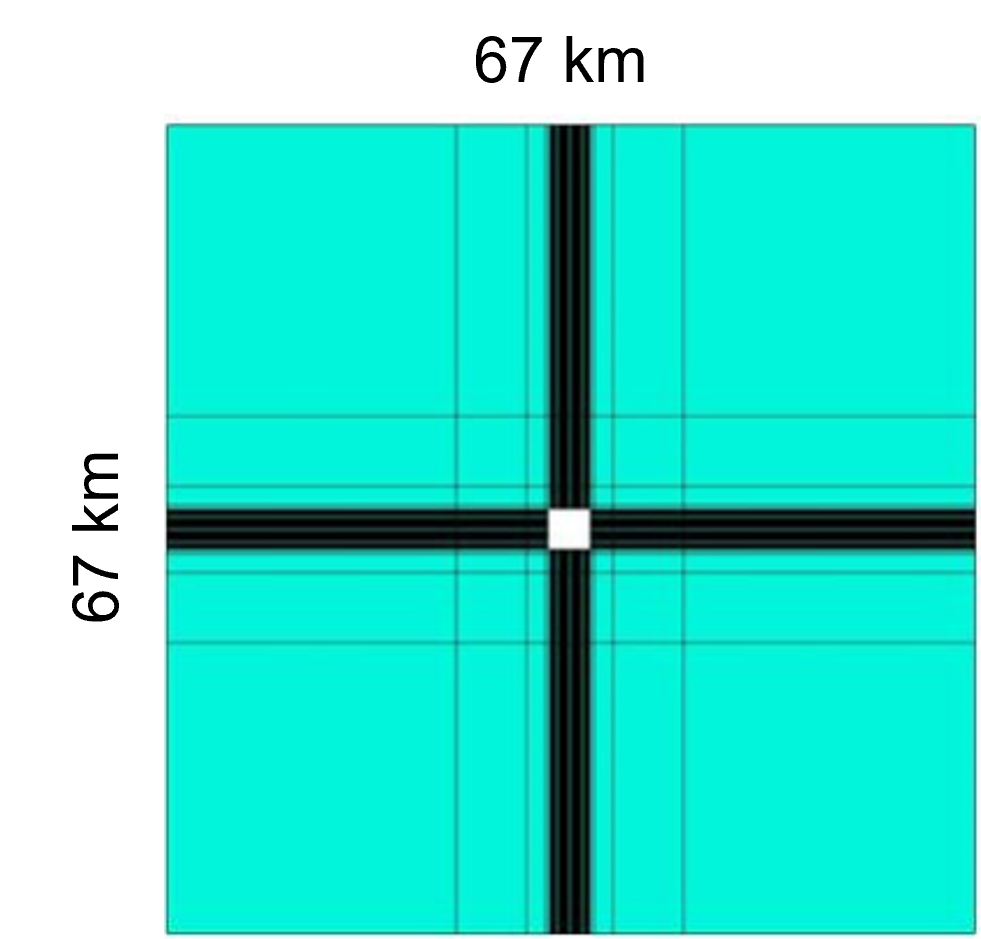}\label{fig:reservoir_model_3d_a}}\hfill
	\subfloat[Equivalent model with the $80 \times 80 \times 20$ storage aquifer (white square) and outer region (in blue)]
 {\includegraphics[width=0.45\textwidth]{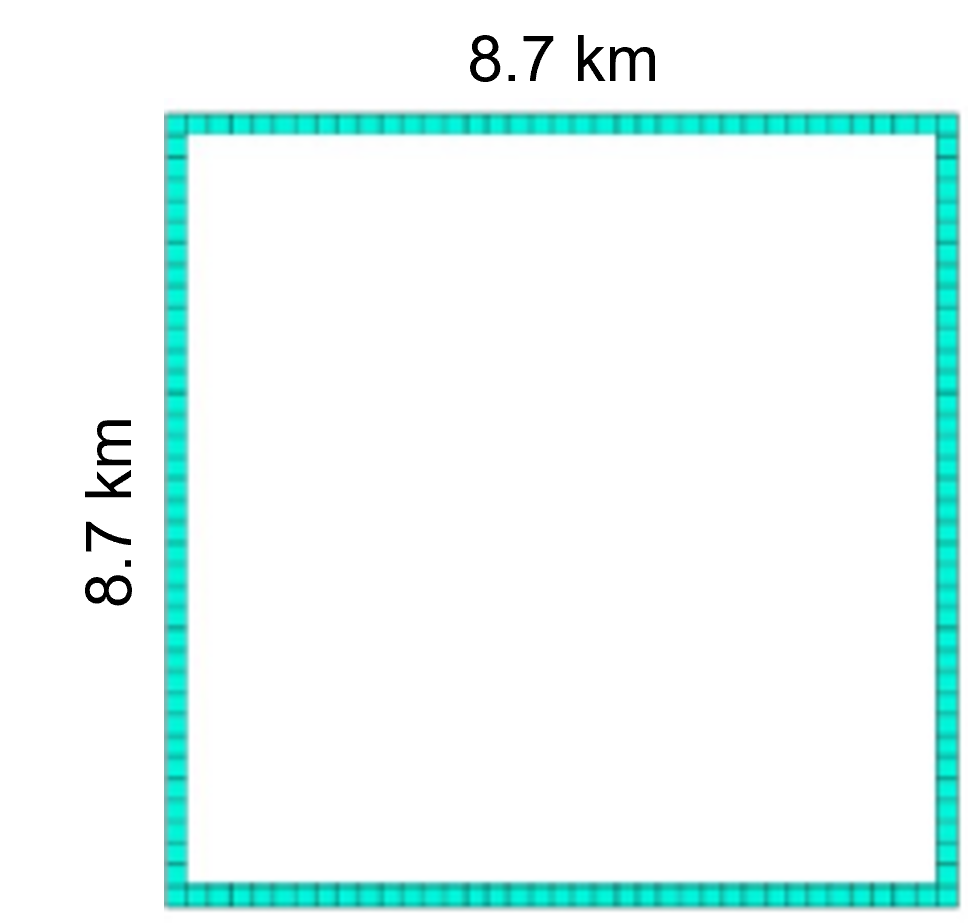}\label{fig:reservoir_model_3d_b}}\hfill
	\subfloat[Log-permeability field for the storage aquifer]{\includegraphics[width=0.6\textwidth]{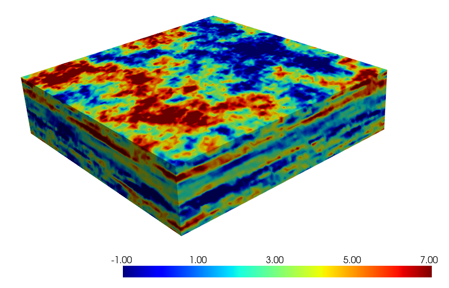}\label{fig:reservoir_model_3d_c}}
    \caption{Overall aquifer models and $\log_e k_x$ ($k_x$ in md) for the storage aquifer.}
	\label{fig:reservoir_model_3d}
\end{figure}

The geological model used here derives from previous carbon storage models developed for the Mt.~Simon formation in central Illinois, USA (Okwen et al.~\citep{okwen2022storage} and Crain et al.~\citep{crain2024integrated}). Our detailed modeling strategy follows that applied by Jiang and Durlofsky~\citep{jiang2024history}. The full model, shown in Fig.~\ref{fig:reservoir_model_3d}, is of size 67~km $\times$ 67~km $\times$ 122~m. It contains a central storage aquifer and an extensive surrounding region to capture large-scale pressure effects. The storage aquifer is 8.5~km $\times$ 8.5~km $\times$ 122~m and is represented by a uniform grid with 80 $\times$ 80 $\times$ 20 cells (a total of 128,000 cells). In the modeling in \citep{jiang2024history}, the grid-block size in the surrounding region increases with distance away from the storage aquifer, as shown in Fig.~\ref{fig:reservoir_model_3d}a. Here, as in \citep{zou2023integrated}, we replace the surrounding region with a single layer of grid blocks that are assigned (very large) pore-volume multipliers such that they have the same total pore volume as the full surrounding region in the original model. The resulting model, of dimensions 82 $\times$ 82 $\times$ 20 cells (a total of 134,480 cells), is shown in Fig.~\ref{fig:reservoir_model_3d}b. Zou and Durlofsky~\citep{zou2023integrated} showed that this treatment provides results that are very close to those using the full model. 

The detailed geomodel for the storage aquifer is depicted in Fig.~\ref{fig:reservoir_model_3d}c. This figure shows $\log_e k_x$, where $k_x=k_y$ are the directional permeabilities in the $x$ and $y$ directions, in md. We take $k_z=0.1 k_x$, where $k_z$ is the permeability in the $z$ direction. 
As in \citep{jiang2024history}, we exclude the uppermost geological layer modeled by Crain et al.~\citep{crain2024integrated}, which is a shale layer with no impact on flow. Our model thus involves the other five geological layers considered in \citep{crain2024integrated}. We use the geostatistical parameters provided in Table~1 in Crain et al.~\citep{crain2024integrated} to generate the geomodel. This treatment differs from that in \citep{jiang2024history}, where a  range of uncertain scenario parameters were considered. 

\begin{figure*}[!htb]
\centering
\includegraphics[width = 0.7\textwidth]{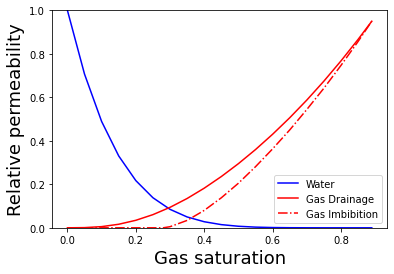}
\caption{Gas-water relative permeability curves used in this study.} \label{fig:rel_perm_3d}
\end{figure*}

We use the Eclipse~300 simulator with the CO2STORE option (Schlumberger~\citep{schlumberger2014eclipse}) for the flow simulations required for GNSM training and testing. The simulation models include two fluid phases (water/brine and gas) and three components (water, CO$_2$, and salt). The initial pressure at the top layer of the model is 155~bar. The system is isothermal, with the temperature specified to be 55$^\circ$C. No-flow boundary conditions are applied on all faces of the overall model. The relative permeability curves, which include the effects of hysteresis, are shown in Fig.~\ref{fig:rel_perm_3d}. 

We consider systems containing four horizontal CO$_2$ injection wells. These wells are subject to prescribed geometric constraints (e.g., minimum well-to-well distance), but they can otherwise be arbitrarily placed within the storage aquifer. Example configurations will be shown in Section~\ref{sec:gnsm_3d}. Each well is prescribed to inject 0.5~Mt~CO$_2$/year (thus the total injection is 2~Mt~CO$_2$/year). Injection proceeds for 20~years, at which time the simulations are terminated. In the GNSM, the goal is to predict the global pressure and saturation fields at $n_t=10$ evenly spaced time steps, i.e., every 2~years. Note there are many more than 10 time steps in the Eclipse CO2STORE simulations.


\section{GNSM framework}
\label{sec:gnsm_3d}

As in Tang and Durlofsky~\citep{tang2024graph}, we construct separate graph neural networks to provide the state variables in all grid blocks at a set of time steps. These are referred to as the pressure graph neural network (PresGNN) and the saturation graph neural network (SatGNN). The PresGNN and SatGNN are trained independently; in combination they comprise the overall GNSM. The networks take as input pressure and saturation at time step $n$, and they predict pressure and saturation at the next time step, $n+1$. The injection rate into each cell containing a well is also predicted in SatGNN. This autoregressive process, which is repeated as we step from $n+1$ to $n+2$, etc., is referred to as one-step rollout. 

We now describe the GNSM framework for the 3D geological modeling of CO$_2$-brine flow driven by horizontal injection wells. The basic workflow, as well as the online application of the GNSM, are analogous to that described in \citep{tang2024graph}. Our emphasis here is on the differences between this model and the 2D oil-water GNSM. Specifically, we discuss the additional features and hyperparameters along with the modified training strategies required for this case.

\subsection{Model architecture}
\label{sec:model_3d}

\begin{figure*}[!htb]
\centering
\includegraphics[width = 0.9\textwidth]{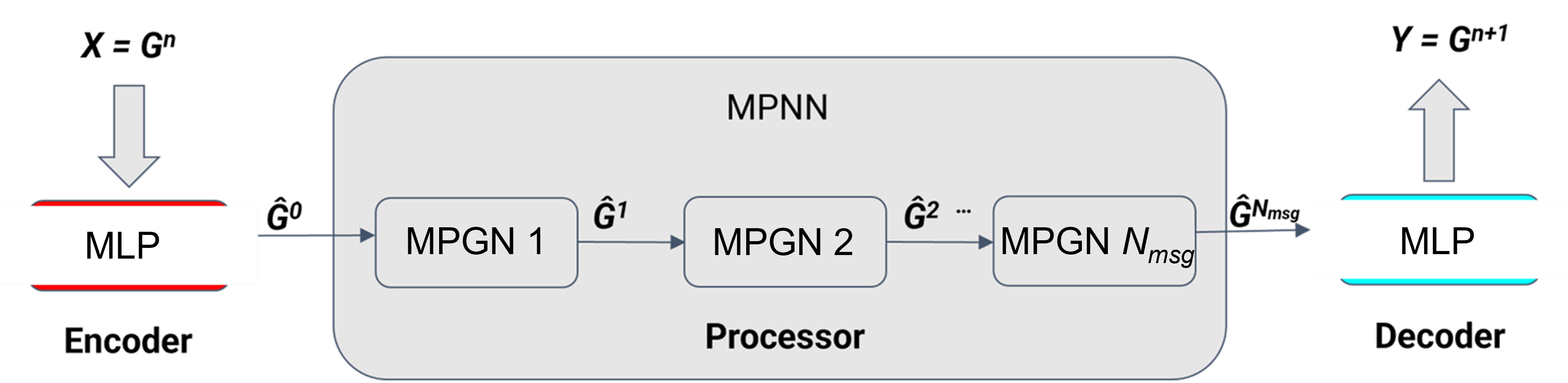}
\caption{Schematic of the model architecture (from~\citep{tang2024graph}). The processor involves a total of $N_{msg}$ message passing neural networks (MPNNs).} \label{fig:model_schematic_3d}
\end{figure*}

Fig.~\ref{fig:model_schematic_3d} provides a schematic representation of PresGNN and SatGNN architectures. The framework consists of three main components: an encoder, a processor, and a decoder. The encoder transforms input features into a latent space, the processor iteratively updates these latent representations by passing messages between neighboring nodes, and the decoder converts the processed latent representations back into the original feature space. This process allows the GNSM to effectively learn and predict complex relationships within the graph data. The processor, which is the main component, involves a message passing neural network (MPNN). The MPNN itself consists of multiple message passing graph networks (MPGNs). The encoder and decoder are multilayer perceptrons (MLPs), used to project data to lower and higher dimensions, respectively.


\begin{figure*}[!htb]
\centering
\includegraphics[width = 0.9\textwidth]{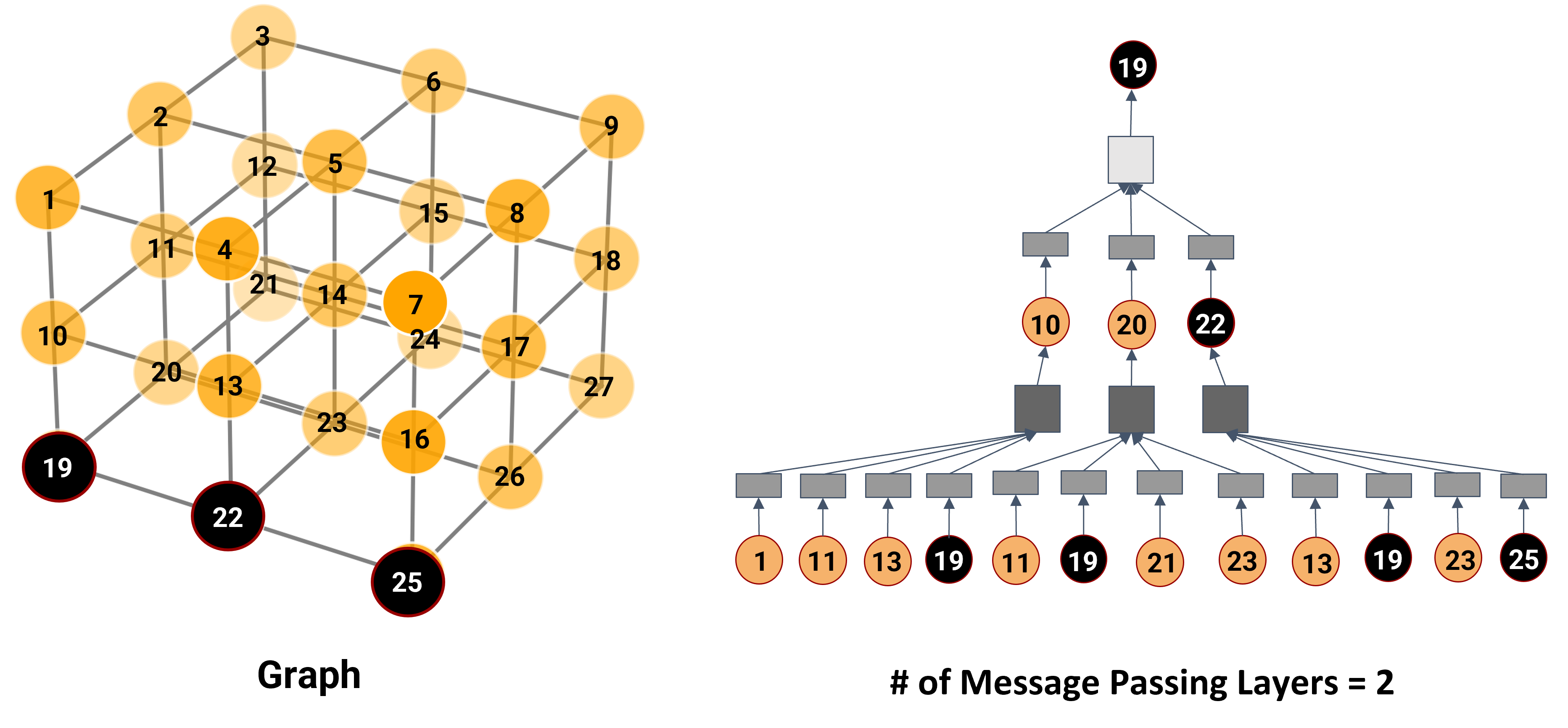}
\caption{MPNN schematic for $3 \times 3 \times 3$ aquifer model (left). Message collection and information aggregation shown on the right for Node~19 with $N_{msg}=2$.} \label{fig:message_passing_3d}
\end{figure*}

A schematic of the MPNN, for an aquifer model containing $3 \times 3 \times 3$ cells with $N_{msg} = 2$, is displayed in Fig.~\ref{fig:message_passing_3d}. Nodes~1--27 in the left image represent grid blocks in the aquifer. Neighboring cells are connected to one another by the graph edges (lines). Node connections are defined such that each node is connected to its immediate neighbors in the $x$, $y$, and $z$ directions (diagonal connections do not appear in the graph). Nodes~19, 22, and 25 (in black) indicate cells that are intersected by a horizontal injection well. Throughout this work, injection wells are assumed to be perforated over their full horizontal extent. The remaining nodes, shown in yellow, correspond to cells that do not contain a well. For each node that will be updated from time level $n$ to $n+1$, we construct a computational graph. This graph is shown, with Node~19 as the target node, in Fig.~\ref{fig:message_passing_3d} (right).

Given that $N_{msg} = 2$ for this example, information is passed across two layers (the direct neighbors and the neighbors of the direct neighbors), as illustrated in Fig.~\ref{fig:message_passing_3d} (right). Two types of networks are involved -- message collection (rectangular boxes) and information aggregation (square boxes). The state for Node~19 is updated using information from a subset of Nodes~1--27 (specifically from its two levels of neighbors). Here $\hat {\mathbf G}^{N_{msg}}$ represents the final encoded graph after $N_{msg}$ steps of the MPGNs, applied to every node. This corresponds to the output from the processor in Fig.~\ref{fig:model_schematic_3d}, which is then fed to the decoder.

\subsection{Features, hyperparameters, and training}
\label{sec:modifications_3d}

The GNSM involves the definition and use of node and edge features. The features used in this study are similar to those considered in~\citep{tang2024graph}, though modifications are required since we are now dealing with 3D models that contain horizontal wells, each intersecting multiple blocks. Node features are listed in Table~\ref{tab:node_features_3d}. In addition to the features considered for the 2D oil-water case, here we include (heterogeneous) cell porosity and cell depth. Well index, which is essentially a transmissibility linking the well to the cell containing it, is computed as in \citep{zou2022effective}.
Injection rate for each well cell, ${Q^n_p}$, also appears in Table~\ref{tab:node_features_3d}. As mentioned earlier, this time-dependent quantity is predicted together with saturation by SatGNN. In contrast to the setup in~\citep{tang2024graph}, we do not use a single-phase pressure solution as a node feature. We have found the current GNSM to provide sufficient accuracy without this information. Finally, we include the current time step length as a feature to account for irregular time step sizes. This feature is not used in the results presented here. 

The edge features are presented in Table~\ref{tab:edge_features_3d}. These involve features related to grid geometry and cell-to-cell transmissibility. Transmissibility can be expressed generically as $k_{f}A/l$, where $k_f$ is an interface permeability, $A$ is the area of the shared face, and $l$ is the distance between the cell centers.

\begin{table}[htb!]
\centering
    \caption{GNSM node features}
    \begin{tabular}{|c|c|c|}
    \hline
    {No.} & {Feature} & {Quantity} \\ \hline
    {1.} & {${p}^n$} & {current pressure}\\ 
    {2.} & {${S}^n_w$} & {current water saturation}\\ 
    {3.} & {${k}$} & {permeability}\\
    {4.} & {${\phi}$} & {porosity}\\
    {5.} & {${V}$} & {cell bulk volume}\\
    {6.} & {${D}$} & {cell depth}\\
    {7.} & {${W}$} & {well index (for well cells)}\\ 
    {8.} & {${Q^n_p}$} & {perforation injection rate (for well cells)}\\ 
    {9.} & {${e}$} & {encoding of node type}\\ 
    {10.} & {${\delta t}^n$} & {current time step length}\\
    \hline
    \end{tabular}%
    \label{tab:node_features_3d}%
\end{table}%

\begin{table}[htb!]
\centering
    \caption{GNSM edge features}
    \begin{tabular}{|c|c|c|}
    \hline
    {No.} & {Feature} & {Quantity} \\ \hline
    {1.} & {${T}$} & {transmissibility}\\ 
    {2.} & {${d}_x$} & {distance in $x$ dimension}\\ 
    {3.} & {${d}_y$} & {distance in $y$ dimension}\\ 
    {4.} & {${d}_z$} & {distance in $z$ dimension}\\ 
    {5.} & {${d}_t$} & {total distance between cells}\\
    \hline
    \end{tabular}%
    \label{tab:edge_features_3d}%
\end{table}%

The hyperparameters used in the GNSM, along with the range considered for each hyperparameter, appear in Table~\ref{tab:hyperparameters_3d}. The ranges for some hyperparameters differ from those in~\citep{tang2024graph} due to the very different model sizes in the two studies (in the 2D oil-water case the models contained 6045~cells, while the models here have 134,480~cells). In addition, two new hyperparameters are introduced -- the first-step loss weight and the CO$_2$ plume loss weight. The first-step loss weight allows us to add extra weight to the initial time step. These states can be more challenging to predict because the system shifts from a static state to a dynamic state over this time step. The CO$_2$ plume loss weight adds extra weight for blocks containing the plume. This leads to better accuracy in predictions of the plume location.

\begin{table}[htb!]
\centering
    \caption{GNSM hyperparameters}
    \begin{tabular}{|c|c|c|}
    \hline
    {No.} & {Hyperparameter} & {Range considered} \\ \hline
    {1.} & {number of hidden layers} & {$2$ or $3$}\\ 
    {2.} & {hidden size} & {$64$ or $128$}\\ 
    {3.} & {number of message passing layers} & {$7$ -- $15$}\\ 
    {4.} & {type of message passing} & {use difference and edge features or not}\\
    {5.} & {type of aggregation function} & {summation, maximization, or averaging}\\
    {6.} & {latent size} & {$16$ or $32$}\\
    {7.} & {types of activation} & {ReLU, Leaky ReLU, or ELU}\\
    {8.} & {group normalization} & {None, Processor, or MLP}\\
    {9.} & {Gaussian noise std.~dev.}   & {$0.01$, $0.03$, or $0.05$}\\ 
    {10.} & {MAE loss ratio} & {$0.1$ -- $0.9$}\\ 
    {11.} & {well loss ratio} & {$0.1$, $1.0$, or $10.0$}\\ 
    {12.} & {number of multistep training} & {1 -- 2}\\
    {13.} & {learning rate} & {$10^{-3}$ -- $10^{-5}$}\\
    {14.} & {first-step loss weight} & {0.1 -- 0.9}\\
    {15.} & {CO$_2$ plume loss weight} & {0.1 -- 0.9}\\
    \hline
    \end{tabular}%
    \label{tab:hyperparameters_3d}%
\end{table}%

\begin{figure}[H]
	\centering
	\subfloat[Well configuration 1]{\includegraphics[width=0.45\textwidth]{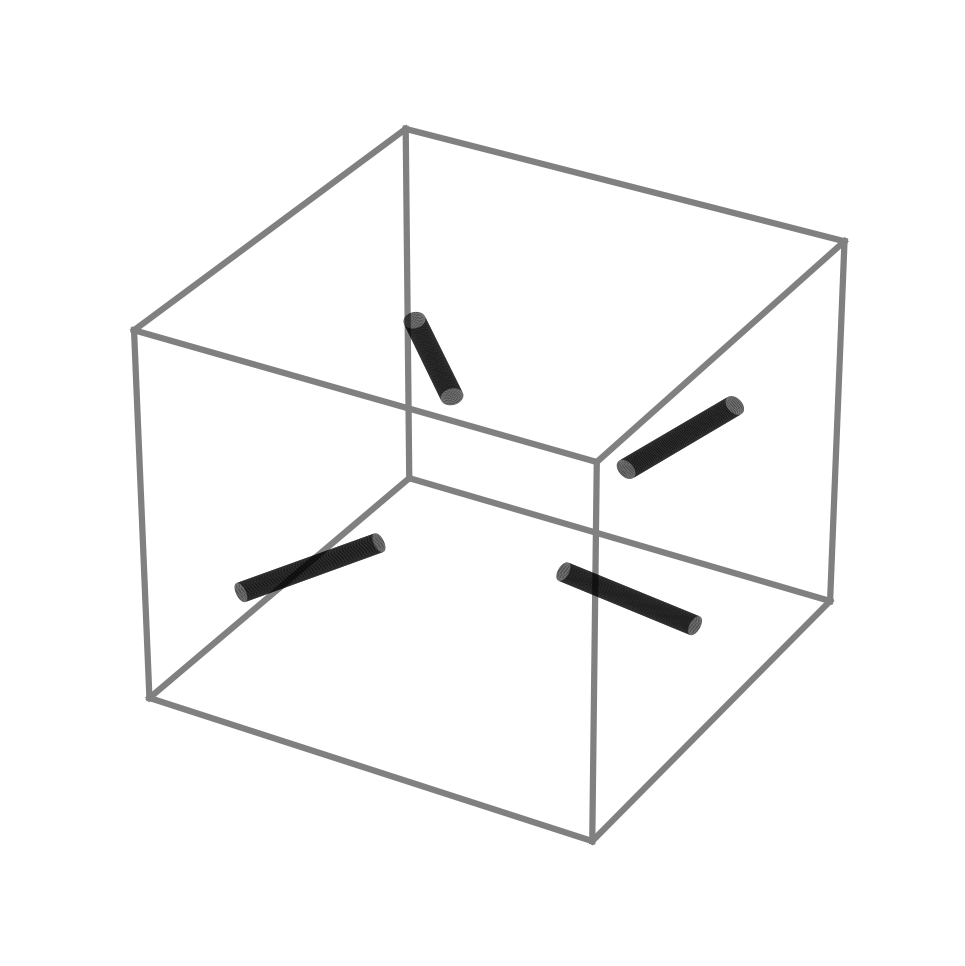}}\hfill
	\subfloat[Well configuration 2]{\includegraphics[width=0.45\textwidth]{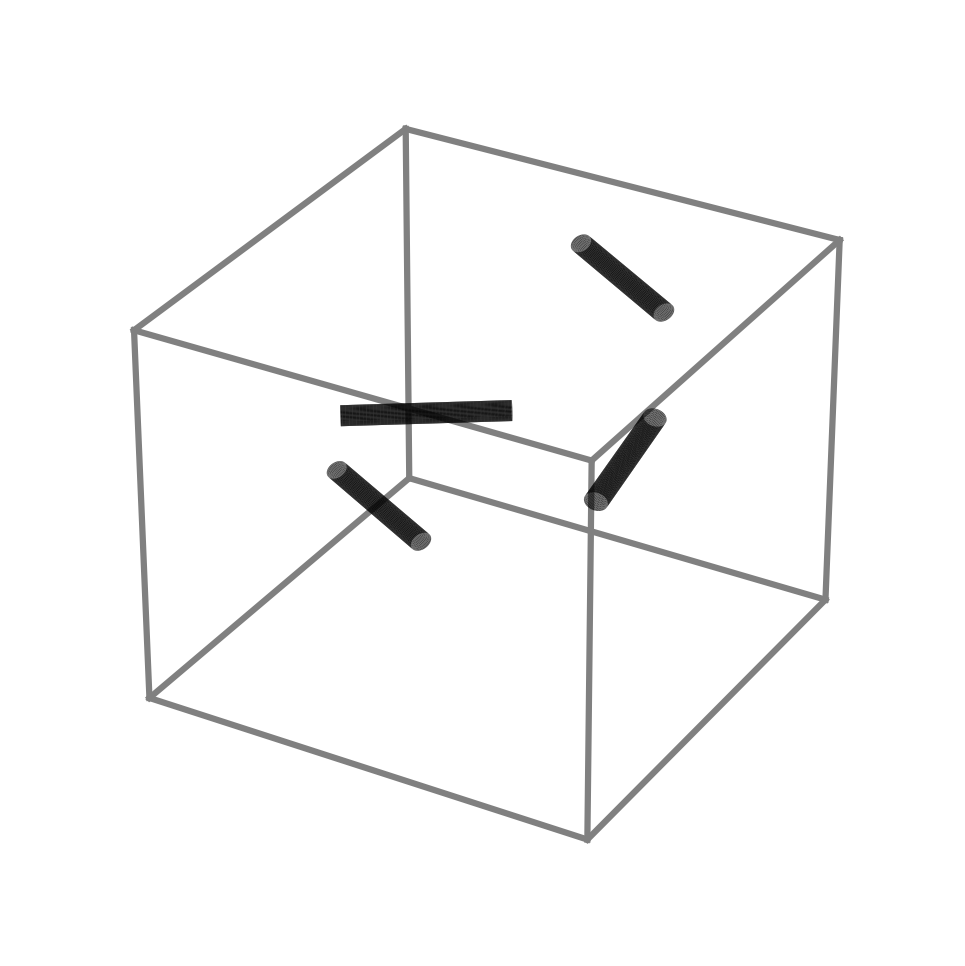}}\hfill
	\subfloat[Well Configuration 3]{\includegraphics[width=0.45\textwidth]{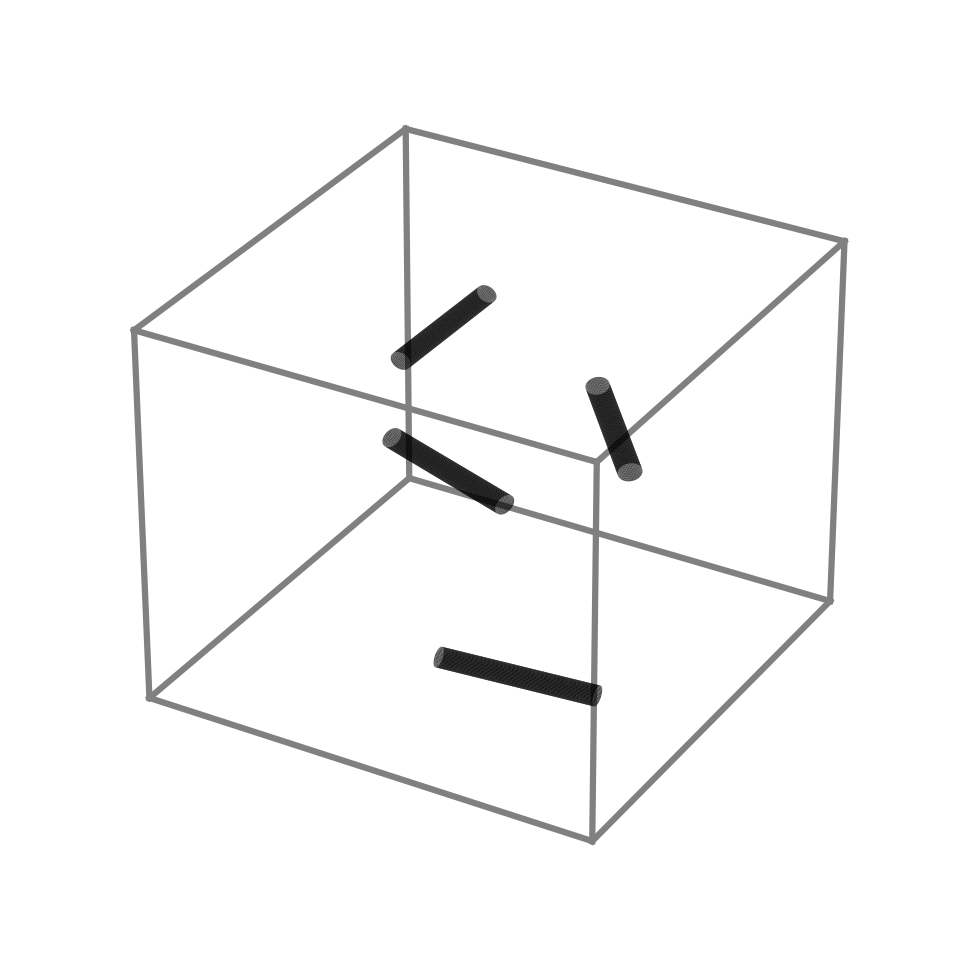}}\hfill
	\subfloat[Well configuration 4]{\includegraphics[width=0.45\textwidth]{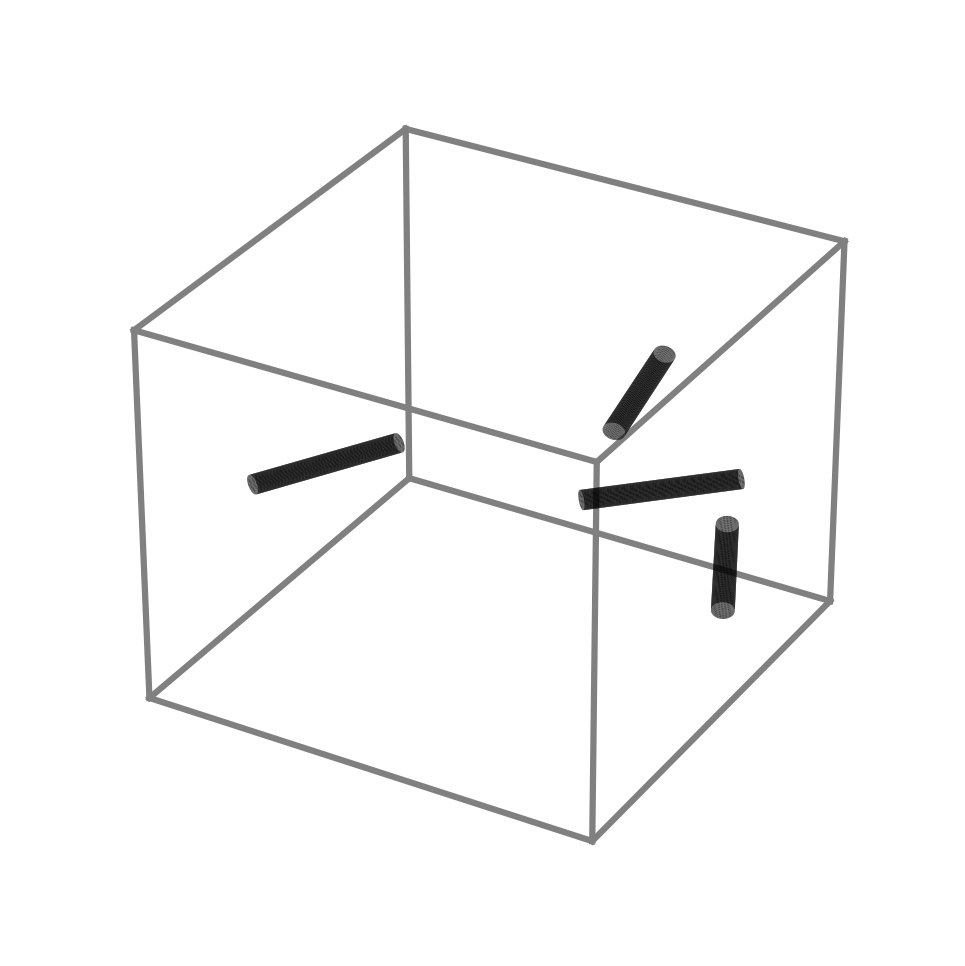}}\hfill
    \caption{Example well configurations used for training and testing. All configurations contain four horizontal injection wells.}
	\label{fig:setup_example_3d}
\end{figure}

For training, we perform 100 flow simulations using Eclipse~300 CO2STORE. The simulation setup is as described in Section~\ref{sec:problem_setup_3d}. In each of these runs, four horizontal wells are placed randomly within the storage aquifer. We impose the same constraints on training and testing configurations as will be applied in the optimizations (e.g., well lengths must be between 480~m and 1200~m). The full set of constraints will be given in Section~\ref{sec:result_opt_3d}. Four random well configurations, of the type used for training and testing, are shown in Fig.~\ref{fig:setup_example_3d}.


We use a multistage approach for training, similar to the strategy introduced in~\citep{tang2024graph}. In the first stage, a wide range of hyperparameters are evaluated using small datasets. Specifically, 10~samples, with each sample representing the simulation results for a particular well configuration, are considered in this stage. In the second stage, we further evaluate the promising hyperparameter sets identified in the first stage. The second stage involves training for a single time step with all 100~samples, with each model trained for 400~epochs with a learning rate of $10^{-3}$. In the final stage we apply multistep training. This approach acts to reduce error accumulation over multiple time steps. Because the total number of GNSM time steps is only 10, we consider up to two additional steps for training. The learning rate in this stage is $3 \times 10^{-5}$, and the models are trained for 100~epochs each.

Consistent with the approach in~\citep{tang2024graph}, we add Gaussian noise, and apply residual prediction, to enhance the robustness of the GNSM. Gaussian noise, with a mean of 0 and a standard deviation specified by hyperparameter~9 in Table~\ref{tab:hyperparameters_3d}, is added to the normalized training data. Specifically, the training data are perturbed as ${\bf {\tilde P}}^n = {\bf P}^{n} + {\bf \Psi}_p$ and ${\bf \tilde{S}}_w^n = {\bf S}^n_w + {\bf \Psi}_s$, where ${\bf P}^n \in \mathbb{R}^{n_c}$ and ${\bf S}_w^n \in \mathbb{R}^{n_c}$ are the normalized simulation data (pressure data are normalized using the overall maximum and minimum pressure, saturation data do not require normalization) and $n_c$ denotes the number of cells in the model. The added Gaussian noise is given by ${\bf \Psi}_p \sim \mathcal{N}(0,\,\sigma_p^{2})$ and ${\bf \Psi}_s \sim \mathcal{N}(0,\,\sigma_s^{2})$, with $\sigma_p$ and $\sigma_s$ as the standard deviations. Residual prediction means the changes in state variables between time steps are predicted, rather than the state variables themselves. We thus predict $\Delta {\bf {\tilde P}}^n = {\bf P}^{n+1}-{\bf {\tilde P}}^{n}$ and $\Delta {\bf \tilde{S}}_w^n = {\bf S}_w^{n+1}-{\bf \tilde{S}}_w^{n}$.  

The loss function ($L$) is given by
\begin{equation}
    \begin{split}
    L =  &\frac{1}{n_{s} n_t} \left( \sum_{i=1}^{n_{s}}\sum_{k=1}^{n_t} \left\| \Delta \hat{\tilde{\mathbf{X}}}^{k}_{v,i} - \Delta \tilde{\mathbf{X}}^k_{v,i} \right\|_{2}^2 + \alpha \sum_{i=1}^{n_{s}}\sum_{k=1}^{n_t} \left\| \Delta \hat{\tilde{\mathbf{X}}}^{k}_{v,i} - \Delta \tilde{\mathbf{X}}^k_{v,i} \right\|_{1} \right) +  \\  
    &\gamma \frac{1}{n_{s} n_t n_c^w} \left( \sum_{i=1}^{n_{s}}\sum_{k=1}^{n_t}\sum_{j=1}^{n_c^w} \left\| \delta \hat{\tilde{x}}^{k,j}_{v,i} - \delta \tilde{x}^{k,j}_{v,i} \right\|_2^2 + \beta \sum_{i=1}^{n_{s}}\sum_{k=1}^{n_t}\sum_{j=1}^{n_c^w} \left\| \delta \hat{\tilde{x}}^{k,j}_{v,i} - \delta \tilde{x}^{k,j}_{v,i} \right\|_1 \right)  + \\   
    &\eta\frac{1}{n_{s}}  \sum_{i=1}^{n_{s}} \left\| \Delta \hat{\tilde{\mathbf{X}}}^{1}_{v,i} - \Delta \tilde{\mathbf{X}}^1_{v,i} \right\|_{2}^2 + \zeta\frac{1}{n_{s} n_t n_p} \sum_{i=1}^{n_{s}}\sum_{k=1}^{n_t}\sum_{l=1}^{n_p} \left\| \delta \hat{\tilde{x}}^{k,l}_{v,i} - \delta \tilde{x}^{k,l}_{v,i} \right\|_2^2.
    \end{split}\label{eq:loss_function_3d}
\end{equation}
Here, $n_s$, $n_t$, $n_c^w$, and $n_p$ are the number of samples (which vary from stage to stage), number of GNSM time steps ($n_t=10$), the total number of cells intersected by all wells (or total number of well completions/perforations), and the number of cells corresponding to the plume, respectively. The parameter $\alpha$ denotes the weight of the mean absolute error (MAE) loss for cells not containing wells, $\beta$ is the weight of the MAE loss for well cells, $\gamma$ indicates the extra weight for well-cell loss (hyperparameter~11), $\eta$ is the weight for first-step loss (hyperparameter~15), and $\zeta$ denotes the CO$_2$ plume loss weight (hyperparameter~16). 

In Eq.~\ref{eq:loss_function_3d}, the vector $\Delta {\bf \hat{\tilde X}^\text{$k$}_\text{$v,i$}}$ is the GNSM residual (perturbed data) prediction for state variable $v$ ($v = p, s$) for sample $i$ at time step $k$, and $\Delta {\bf {\tilde X}}^k_{v,i}$ is its simulation counterpart (also perturbed data). The scalars ${\delta \hat{\tilde x}}^{k,j}_{v,i}$ and ${\delta {\tilde x}}^{k,j}_{v,i}$ denote the same quantities for well blocks, and the scalars ${\delta \hat{\tilde x}}^{k,l}_{v,i}$ and ${\delta {\tilde x}}^{k,l}_{v,i}$ are analogous variables for cells that are within the CO$_2$ plume. These are taken to be cells for which $S_g > 0.1$, where $S_g$ is CO$_2$ saturation. Within SatGNN, we also predict ${Q^n_p}$, the injection rate into each well block. The scalars ${\delta \hat{\tilde x}}^{k,j}_{s,i}$ and ${\delta {\tilde x}}^{k,j}_{s,i}$ (for $v = s$) thus contain injection rate, in addition to saturation, at well blocks.

\begin{table}[htb!]
\centering
    \caption{Optimized GNSM hyperparameters}
    \begin{tabular}{|c|c|c|c|}
    \hline
    {No.} & {Hyperparameter} & {PresGNN} & {SatGNN} \\ \hline
    {1.} & {number of hidden layers} & {$2$} & {$2$}\\ 
    {2.} & {hidden size} & {$128$} & {$128$}\\ 
    {3.} & {number of message passing layers} & {$15$} & {$10$}\\ 
    {4.} & {type of message passing} & {all info} & {all info}\\
    {5.} & {type of aggregation function} & {summation} & {summation}\\
    {6.} & {latent size} & {$32$} & {$32$}\\
    {7.} & {types of activation} & {LeakyReLU} & {LeakyReLU}\\
    {8.} & {group normalization} & {None} & {MLP}\\
    {9.} & {Gaussian noise std.~dev.}   & {$0.03$}   & {$0.01$}\\ 
    {10.} & {MAE loss ratio} & {$0.1$} & {$0.1$}\\ 
    {11.} & {well loss ratio} & {$0.1$} & {$0.1$}\\ 
    {12.} & {number of multistep training} & {2} & {2}\\
    {13.} & {learning rate (stage~1~and~2)} & {$10^{-3}$} & {$10^{-3}$}\\
    {14.} & {learning rate (stage~3)} & {$3 \times 10^{-5}$} & {$3 \times 10^{-5}$}\\
    {15.} & {first-step loss weight} & {$0.3$} & {$0.3$}\\
    {16.} & {CO$_2$ plume loss weight} & {$0.3$} & {$0.3$}\\    
    \hline
    \end{tabular}%
    \label{tab:opt_hyperparameters_3d}%
\end{table}%

The final optimized hyperparameters for the PresGNN and SatGNN are given in Table~\ref{tab:opt_hyperparameters_3d}. 
Training each network, with a batch size of~4, takes about 50~hours on a single Nvidia A100 GPU (without parallelization). Although training time increases for models with more cells/nodes, the number of required training samples decreases. Specifically, for the oil-water case with 6045~cells, 600 training samples were used, and 30~hours were required to train each network. The need for fewer training samples here is because GNNs learn from local patterns within data samples, and larger models provide more local patterns, so fewer training samples are needed. This observation is consistent with the findings of Wu et al.~\citep{wu2022learning}, who used only 20 samples to train a GNN surrogate for a model with $\sim10^6$ cells.

\subsection{Bottom-hole pressure prediction}
\label{sec:bhp}

Bottom-hole pressures (BHPs) are important quantities in carbon storage and other reservoir engineering applications. Wells can be controlled by specifying BHP rather than rate, and BHPs often appear as constraints in simulation and optimization problems. Thus the accurate prediction of BHP is required within our framework. In this study, we train a single multilayer perceptron (MLP) to predict BHPs using the node features shown in Table~\ref{tab:node_features_3d}. The same dataset used for training the GNSM is used for this training, so no additional flow simulation is entailed.

The features that are input to the MLP for a particular horizontal well include pressure, saturation, injection rates, etc., for all cells intersected by the well (recall wells are assumed to be perforated over their full horizontal extent). Because the length of each well can vary within a specified range, different numbers of blocks are intersected from well to well and from case to case. To ensure uniform input data length, we use the maximum possible number of intersected cells as the input size. If a well intersects fewer cells, we  duplicate some of the well-cell information to meet this requirement. In the runtime computations, the input features required by the MLP are provided by the PresGNN and SatGNN models. 

During numerical experimentation, we observed that BHP predictions in the first time step were often noticeably less accurate. To circumvent this problem, we use a data augmentation technique during training, in which first-step data are included four times in the training dataset. This approach was found to significantly improve MLP performance in the first time step, while retaining accuracy in subsequent steps.

The loss function for BHP training, which acts to minimize the mean squared error between the predicted and true BHP values, is given by
\begin{equation}
    \begin{split}
    L = \frac{1}{n_s (n_t+3) n_w} \sum_{i=1}^{n_{s}}\sum_{j=1}^{n_{t}+3}\sum_{k=1}^{n_{w}}\left\| \hat{{x}}_{i}^{j,k} -  {x}_{i}^{j,k} \right\|_2^2 ,
    \end{split}\label{eq:loss_function_bhp_3d}
\end{equation}
where $\hat{{x}}_{i}^{j,k}$ and ${x}_{i}^{j,k}$ are the MLP prediction and simulation result for the normalized BHP of well $k$ in data sample $i$ at time step $j$. Note that the number of wells ($n_w$) appears here (rather than $n_c^w$ as in Eq.~\ref{eq:loss_function_3d}) because there is one BHP per well at a given time step, regardless of how many cells the well intersects. As a result of data augmentation, $n_t+3$ time steps are considered in Eq.~\ref{eq:loss_function_bhp_3d}. The training of this MLP requires around 1~hour on a single Nvidia A100 GPU (without parallelization).

\section{GNSM performance for test cases}
\label{sec:result_test_3d}

We now assess the performance of the GNSM. This evaluation includes aggregate error statistics, visual comparisons of pressure, saturation and BHPs for particular test cases, and an assessment of the GNSM for the prediction of CO$_2$ footprint, i.e., storage efficiency. In terms of timings, an Eclipse simulation run takes about 8~minutes, while a full GNSM evaluation 
(after training) requires approximately 4~seconds, about half of which is overhead. Thus we achieve a runtime speedup factor of 120.

\subsection{Overall error computation for test samples}
\label{sec:result_error_3d}
We consider a test set of $n_e=50$ models. These models involve new (arbitrary, though subject to geometric constraints) well configurations  that were not used in training. We simulate the test set using Eclipse CO2STORE, which provides the reference results against which GNSM predictions will be compared.

We compute the relative errors for pressure ($e_p^i$) and saturation ($e_s^i$) for test sample $i$ ($i=1,\ldots,n_e$), through application of
\begin{equation}\label{eq:rel_err_pressure_sat_3d}
    e_p^i = \frac{1}{n_c n_t} \sum_{j=1}^{n_c} \sum_{t=1}^{n_t} \frac{\left| \hat{P}^t_{i,j} - P^t_{i,j}\right|}{P_{max}-P_{min}}, \ \ \  e_s^i = \frac{1}{n_c n_t} \sum_{j=1}^{n_c} \sum_{t=1}^{n_t} \frac{\left| \hat{S}^t_{i,j} - S^t_{i,j}\right|}{S^t_{i,j}+\epsilon}.
\end{equation}
Here $n_c$ and $n_t$ are the number of cells in the model and the number of time steps. The variables $\hat{P}^t_{i,j}$ and $P^t_{i,j}$ refer to the cell pressure predicted by GNSM and the simulator, respectively, for cell $j$ at time step $t$ in test sample $i$. The values $P_{max}$ and $P_{min}$ are the maximum and minimum pressures in sample $i$ at time step $t$. Similarly, $\hat{S}^t_{i,j}$ and $S^t_{i,j}$ denote the saturation values from the GNSM and the simulator, respectively, for cell $j$ at time step $t$ in test sample $i$. Because the initial gas saturation is 0, we introduce $\epsilon = 0.05$ in the denominator of the $e_s^i$ expression. 

\begin{figure*}[!htb]
\centering
\includegraphics[width = 0.9\textwidth]{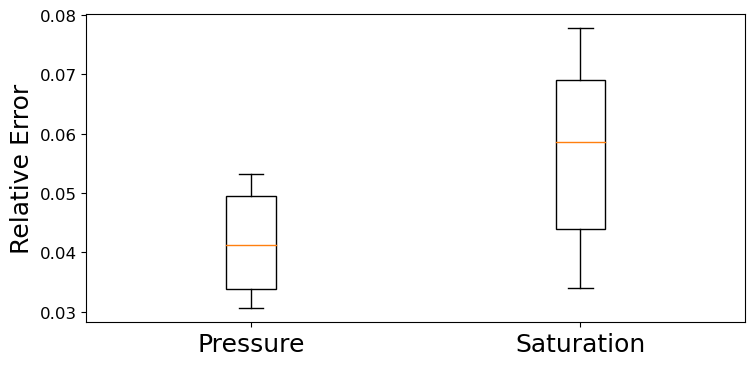}
\caption{Box plots of relative errors for state variables over the 50 test cases. Boxes display P$_{90}$, P$_{75}$, P$_{50}$, P$_{25}$ and P$_{10}$ (percentile) errors. Error calculations given in Eq.~\ref{eq:rel_err_pressure_sat_3d}.} \label{fig:state_map_3d}
\end{figure*}

Fig.~\ref{fig:state_map_3d} shows box plots of the relative errors for pressure and saturation over the 50 test cases. The whiskers extending above and below each box indicate the P$_{90}$ and P$_{10}$ percentile errors, the box boundaries display the P$_{75}$ and P$_{25}$ errors, and the orange line inside each box shows the P$_{50}$ (median) error. The median errors for pressure and saturation are about 4\% and 6\%, respectively, and the P$_{90}$ errors are about 5.5\% and 7.5\%. Errors are somewhat higher here than in the oil-water case where, for example, the P$_{90}$ errors were about 1.5\% and 3\%. The higher errors here are presumably due to the more challenging nature of our 3D setup. In any event, the errors in Fig.~\ref{fig:state_map_3d} are within a reasonable range and suggest that the GNSM will be useful for optimization and other applications.

\subsection{Results for representative test cases}
\label{sec:result_3d}
We now display pressure and saturation fields, at the end of the simulation time frame (20~years), for four different cases. These cases were selected to illustrate a range of CO$_2$ plume patterns. The overall error, calculated as $(e^i_p + e^i_s)/2$, will be provided for each case. BHP profiles will also be presented for these cases.

\begin{figure}[H]
	\centering
	\subfloat[Simulation saturation in layer~1]{\includegraphics[width=0.4\textwidth]{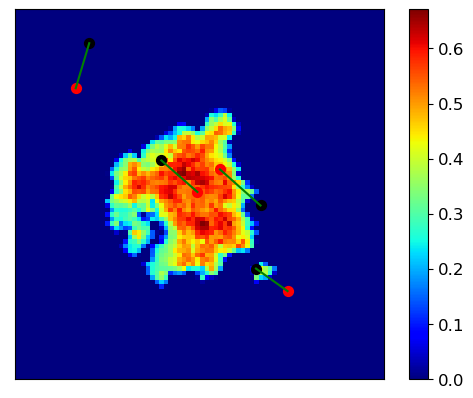}}\hfill
	\subfloat[GNSM saturation in layer~1]{\includegraphics[width=0.4\textwidth]{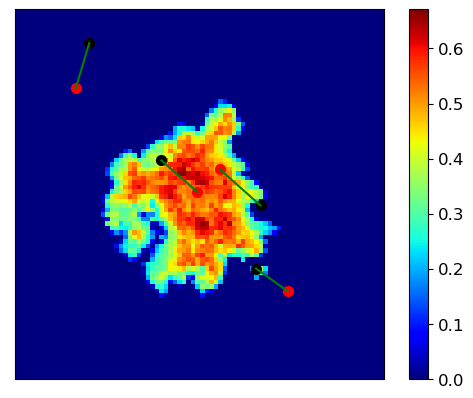}}\hfill
	\subfloat[Simulation saturation in 3D]{\includegraphics[width=0.4\textwidth]{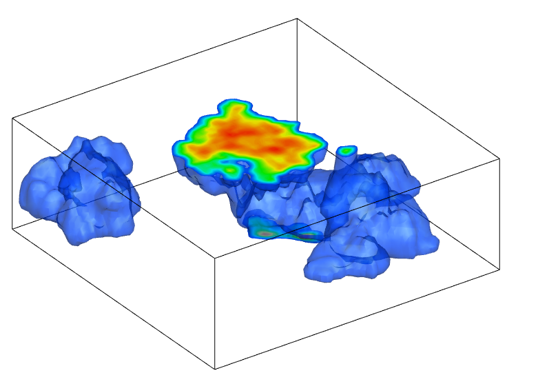}}\hfill
	\subfloat[GNSM saturation in 3D]{\includegraphics[width=0.4\textwidth]{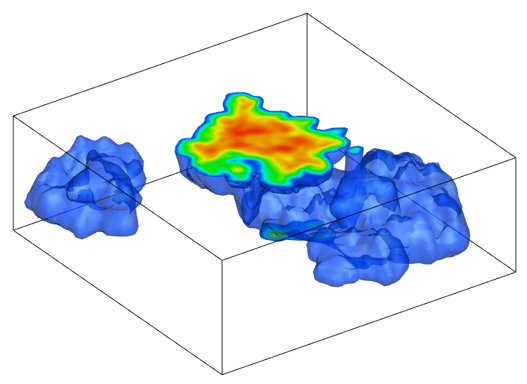}}\hfill
    \caption{Saturation fields for simulation (left) and GNSM (right) for test case~1 at 20~years. Green lines in upper row display the horizontal injectors projected onto the first layer. This case corresponds to P$_{26}$ in overall error.}
	\label{fig:saturation_maps_test1_3d}
\end{figure}

The saturation fields for test case~1 are shown in Fig.~\ref{fig:saturation_maps_test1_3d}. In these images, water appears in blue and gas (CO$_2$) is in red. Test case~1 corresponds to P$_{26}$ in overall error over all test samples, so this is a better-than-average result. This and subsequent figures of this type show reference simulation results on the left and GNSM predictions on the right. The upper row displays results for the first (top) layer in the model, and the lower row provides 3D views. The 3D images display saturation values for cells in which $S_g > 0.05$. The horizontal wells are projected onto the first layer in the 2D saturation maps. The wells appear as green lines, with the heel and toe of each horizontal well indicated by the black and red circles, respectively. For test case~1, we observe close agreement between simulation results and SatGNN predictions. This is evident both in the 3D fields and in the top layer. There are however some minor discrepancies, e.g., the plume is slightly greater in extent in Fig.~\ref{fig:saturation_maps_test1_3d}b than in Fig.~\ref{fig:saturation_maps_test1_3d}a. 

\begin{figure}[H]
	\centering
	\subfloat[Simulation pressure]{\includegraphics[width=0.4\textwidth]{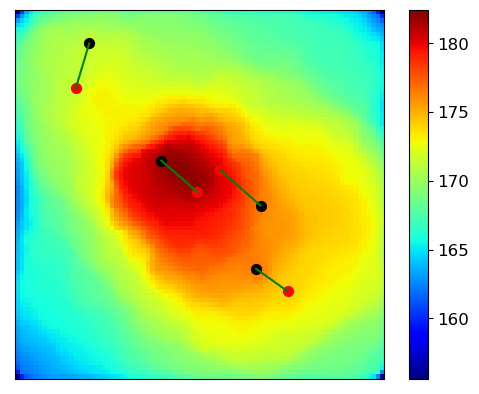}}\hfill
	\subfloat[GNSM pressure]{\includegraphics[width=0.4\textwidth]{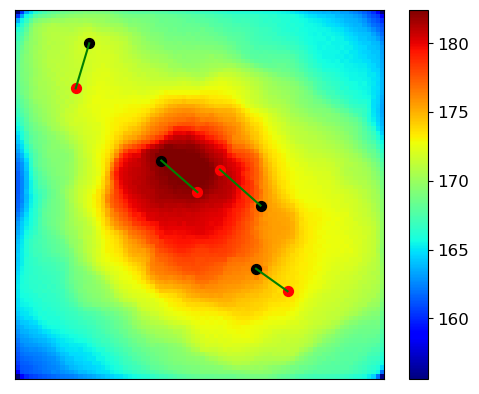}}\hfill
    \caption{Pressure maps in layer~1 for test case~1 at 20~years from simulation (left) and GNSM (right). Green lines depict horizontal injectors projected onto the first layer. This case corresponds to P$_{26}$ in overall error.}
	\label{fig:pressure_maps_and_bhp_test1_3d}
\end{figure}

Pressure maps (pressures are in bar) for the top layer are shown in Fig.~\ref{fig:pressure_maps_and_bhp_test1_3d}. We do not present 3D views for pressure because key features are not as visible for pressure as for saturation (and errors are generally smaller for pressure). It is evident from the figure that the pressure field is captured accurately by PresGNN, though small differences are evident in some locations, such as toward the right edge of the model.

\begin{figure}[H]
	\centering
	\subfloat[Simulation saturation layer~1]{\includegraphics[width=0.4\textwidth]{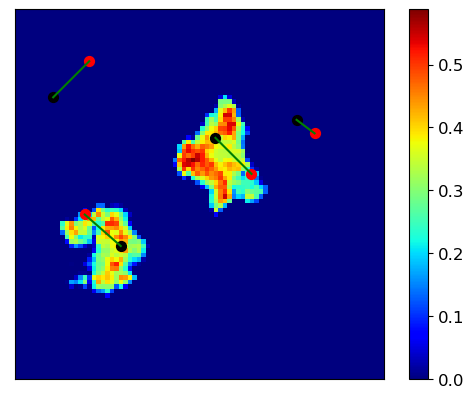}}\hfill
	\subfloat[GNSM saturation in layer~1]{\includegraphics[width=0.4\textwidth]{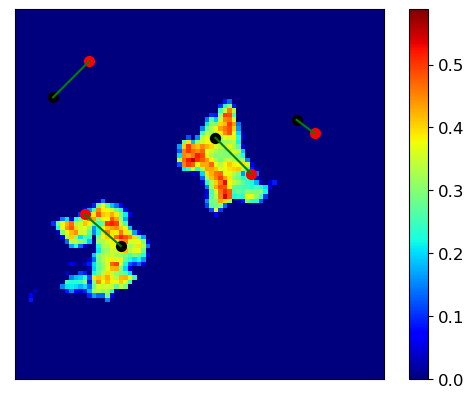}}\hfill
	\subfloat[Simulation saturation in 3D]{\includegraphics[width=0.4\textwidth]{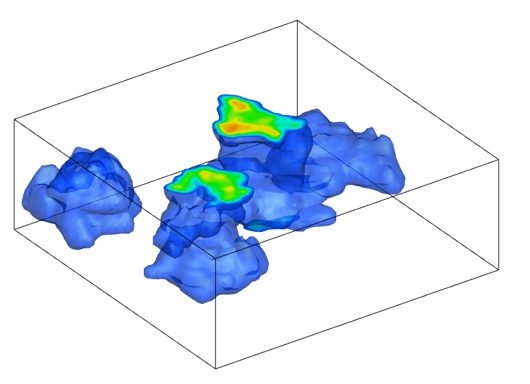}}\hfill
	\subfloat[GNSM saturation in 3D]{\includegraphics[width=0.4\textwidth]{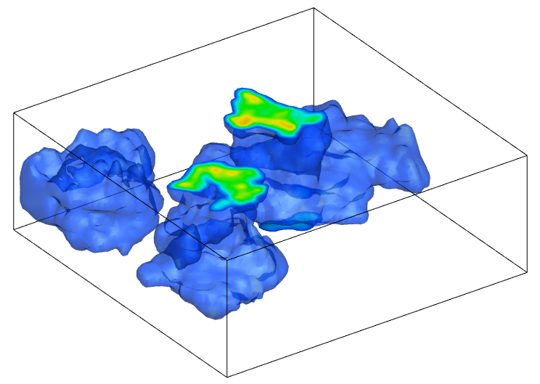}}\hfill
    \caption{Saturation fields for simulation (left) and GNSM (right) for test case~2 at 20~years. Green lines in upper row display the horizontal injectors projected onto the first layer. This case corresponds to P$_{50}$ in overall error.}
	\label{fig:saturation_maps_test2_3d}
\end{figure}

\begin{figure}[H]
	\centering
	\subfloat[Simulation pressure]{\includegraphics[width=0.4\textwidth]{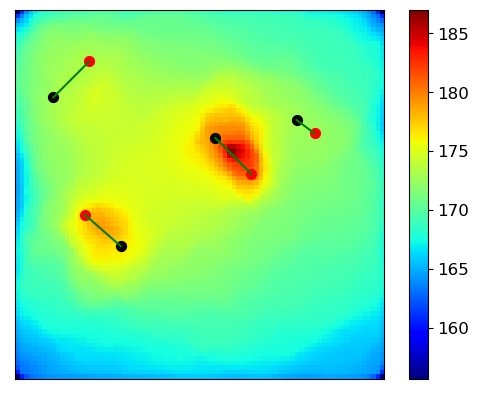}}\hfill
	\subfloat[GNSM pressure]{\includegraphics[width=0.4\textwidth]{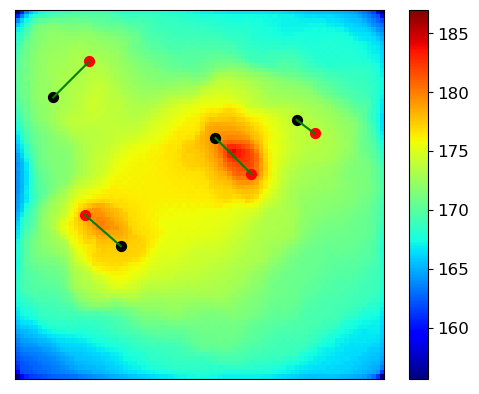}}\hfill
    \caption{Pressure maps in layer~1 for test case~2 at 20~years from simulation (left) and GNSM (right). Green lines depict horizontal injectors projected onto the first layer. This case corresponds to P$_{50}$ in overall error.}
	\label{fig:pressure_maps_and_bhp_test2_3d}
\end{figure}

Test case~2 corresponds to the P$_{50}$ overall state error. Results for this case are presented in Figs.~\ref{fig:saturation_maps_test2_3d} and \ref{fig:pressure_maps_and_bhp_test2_3d}. Here we see two separate CO$_2$ plumes in the top layer of the model, in contrast to the single large plume in the top layer in Fig.~\ref{fig:saturation_maps_test1_3d}. The saturation fields from simulation and SatGNN are again in reasonable agreement, though slight differences in plume shape and size can be seen between Fig.~\ref{fig:saturation_maps_test2_3d}c and d. The pressure fields in Fig.~\ref{fig:pressure_maps_and_bhp_test2_3d} also correspond closely, though there are small differences in the region between the two central (projected) injectors.


\begin{figure}[H]
	\centering
	\subfloat[Simulation saturation in layer~1]{\includegraphics[width=0.4\textwidth]{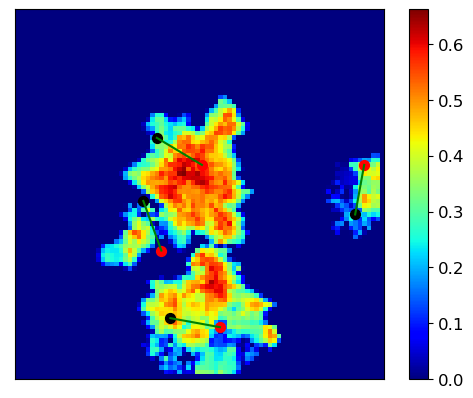}}\hfill
	\subfloat[GNSM saturation in layer~1]{\includegraphics[width=0.4\textwidth]{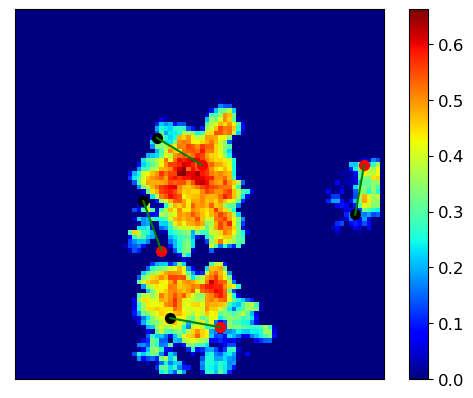}}\hfill
	\subfloat[Simulation saturation in 3D]{\includegraphics[width=0.4\textwidth]{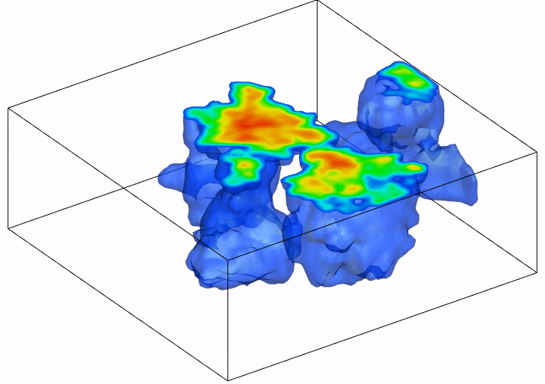}}\hfill
	\subfloat[GNSM saturation in 3D]{\includegraphics[width=0.4\textwidth]{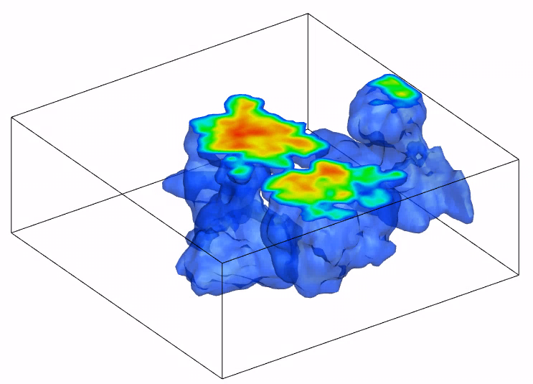}}\hfill
    \caption{Saturation fields for simulation (left) and GNSM (right) for test case~3 at 20~years. Green lines in upper row display the horizontal injectors projected onto the first layer. This case corresponds to P$_{66}$ in overall error.}
	\label{fig:saturation_maps_test3_3d}
\end{figure}

\begin{figure}[H]
	\centering
	\subfloat[Simulation pressure]{\includegraphics[width=0.4\textwidth]{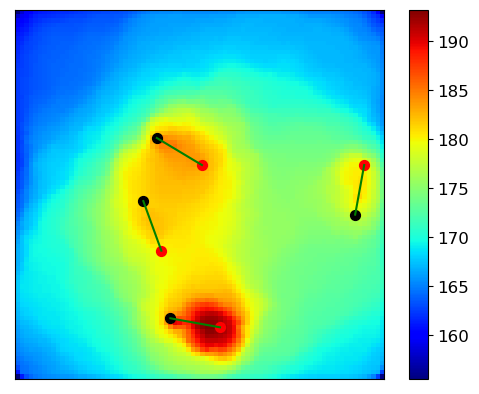}}\hfill
	\subfloat[GNSM pressure]{\includegraphics[width=0.4\textwidth]{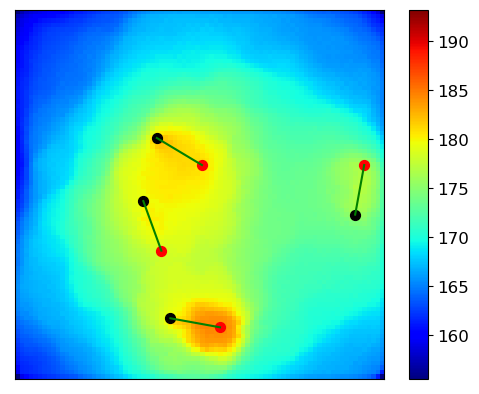}}\hfill
    \caption{Pressure maps in layer~1 for test case~3 at 20~years from simulation (left) and GNSM (right). Green lines depict horizontal injectors projected onto the first layer. This case corresponds to P$_{66}$ in overall error.}
	\label{fig:pressure_maps_and_bhp_test3_3d}
\end{figure}

Results for test case~3, which corresponds to P$_{66}$ in overall error, appear in Figs.~\ref{fig:saturation_maps_test3_3d} and~\ref{fig:pressure_maps_and_bhp_test3_3d}. Here we see saturation fields that are quite different from those in test cases~1 and 2, with CO$_2$ in multiple plumes in the top layer. Despite the complicated plume shapes, SatGNN continues to provide reasonably accurate predictions. Note that in this case some of the CO$_2$ leaves the storage aquifer. Such cases will occur for some well configurations during optimization and thus need to be handled by GNSM. More error is visible in the pressure field for this case than in the previous examples. Specifically, pressure appears to be underpredicted (by several bar) in the vicinity of the projected wells.


\begin{figure}[H]
	\centering
	\subfloat[Simulation saturation in layer~1]{\includegraphics[width=0.4\textwidth]{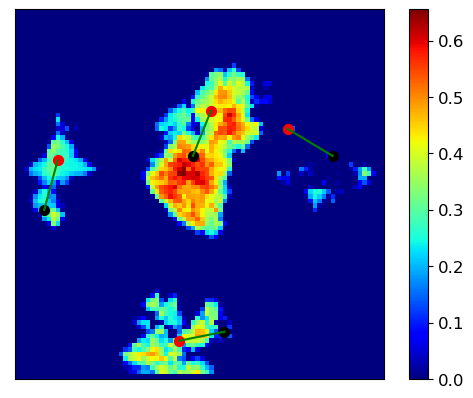}}\hfill
	\subfloat[GNSM saturation in layer~1]{\includegraphics[width=0.4\textwidth]{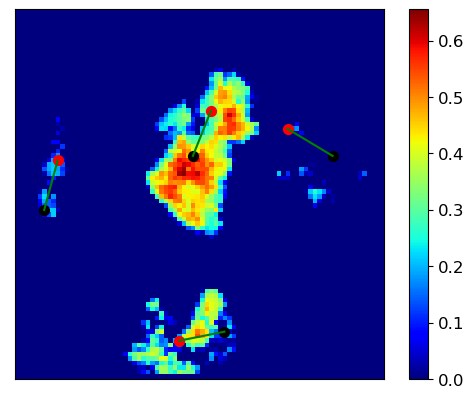}}\hfill
	\subfloat[Simulation saturation in 3D]{\includegraphics[width=0.4\textwidth]{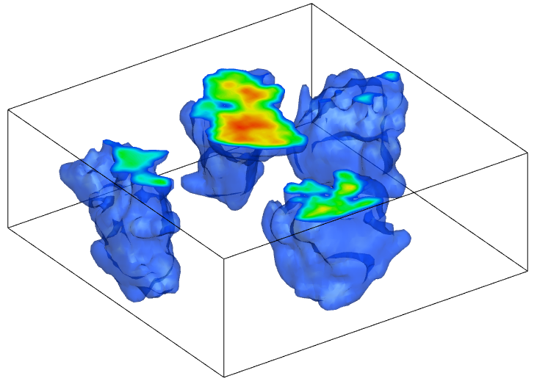}}\hfill
	\subfloat[GNSM saturation in 3D]{\includegraphics[width=0.4\textwidth]{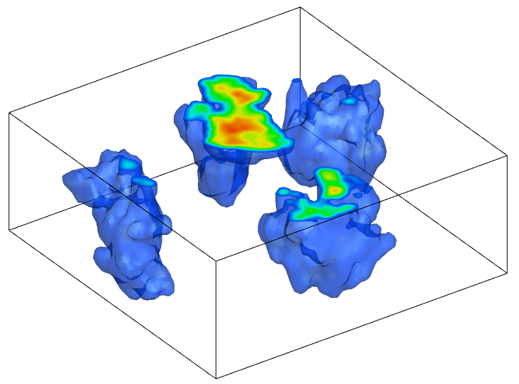}}\hfill
    \caption{Saturation fields for simulation (left) and GNSM (right) for test case~4 at 20~years. Green lines in upper row display the horizontal injectors projected onto the first layer. This case corresponds to P$_{68}$ in overall error.}
	\label{fig:saturation_maps_test4_3d}
\end{figure}

\begin{figure}[H]
	\centering
	\subfloat[Simulation pressure]{\includegraphics[width=0.4\textwidth]{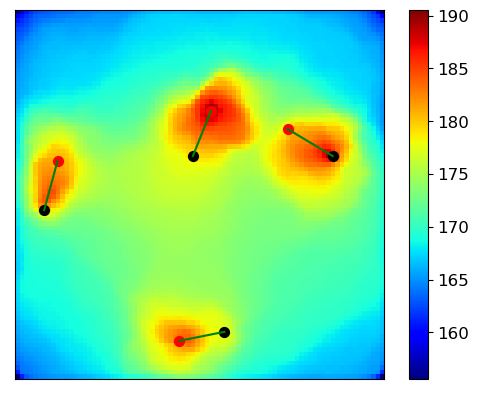}}\hfill
	\subfloat[GNSM pressure]{\includegraphics[width=0.4\textwidth]{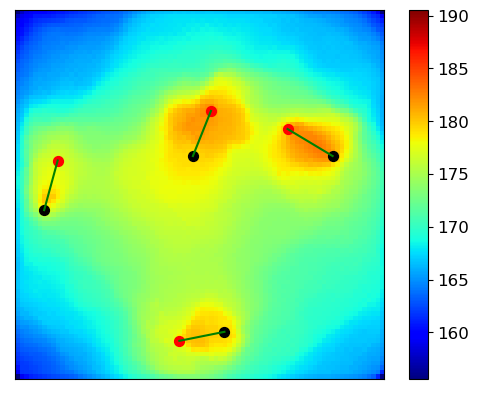}}\hfill
    \caption{Pressure maps in layer~1 for test case~4 at 20~years from simulation (left) and GNSM (right). Green lines depict horizontal injectors projected onto the first layer. This case corresponds to P$_{68}$ in overall error.}
	\label{fig:pressure_maps_and_bhp_test4_3d}
\end{figure}

Test case~4 results are displayed in Figs.~\ref{fig:saturation_maps_test4_3d} and \ref{fig:pressure_maps_and_bhp_test4_3d}. In this case, which corresponds to P$_{68}$ in overall error, CO$_2$ appears in multiple locations in layer~1. The saturation distribution is generally captured by SatGNN, but there are clear differences between SatGNN and the reference simulation results. The extent of the plume on the left of the model, for example, is underpredicted in Fig.~\ref{fig:saturation_maps_test4_3d}b relative to Fig.~\ref{fig:saturation_maps_test4_3d}a. In addition, the PresGNN result underpredicts near-well pressure, as was also observed for test case~3. In an overall sense, however, the GNSM predictions clearly resemble the simulation results.


\begin{figure}[H]
	\centering
	\subfloat[BHP for test case~1]{\includegraphics[width=0.9\textwidth]{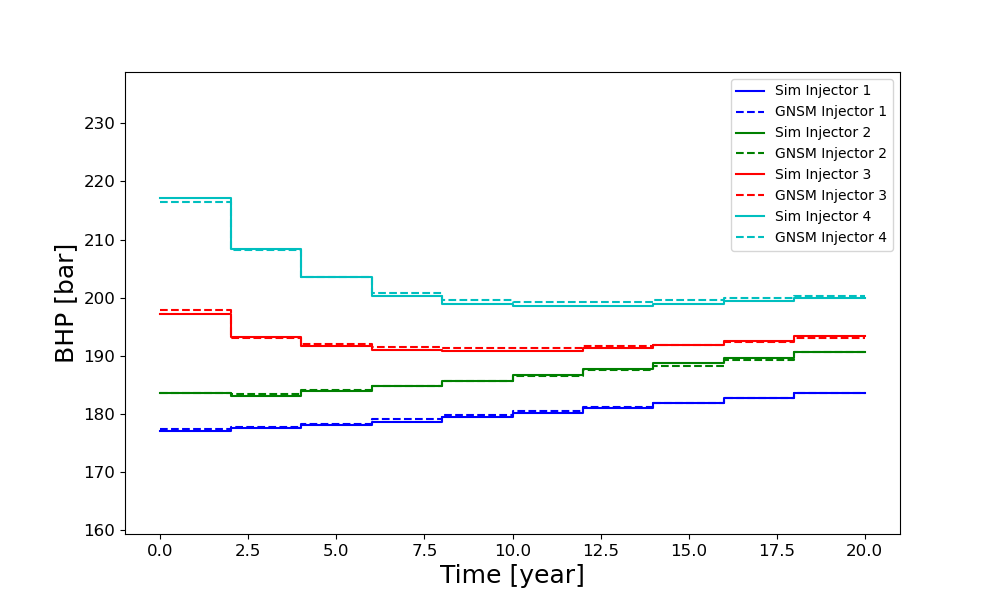}}\hfill
	\subfloat[BHP for test case~2]{\includegraphics[width=0.9\textwidth]{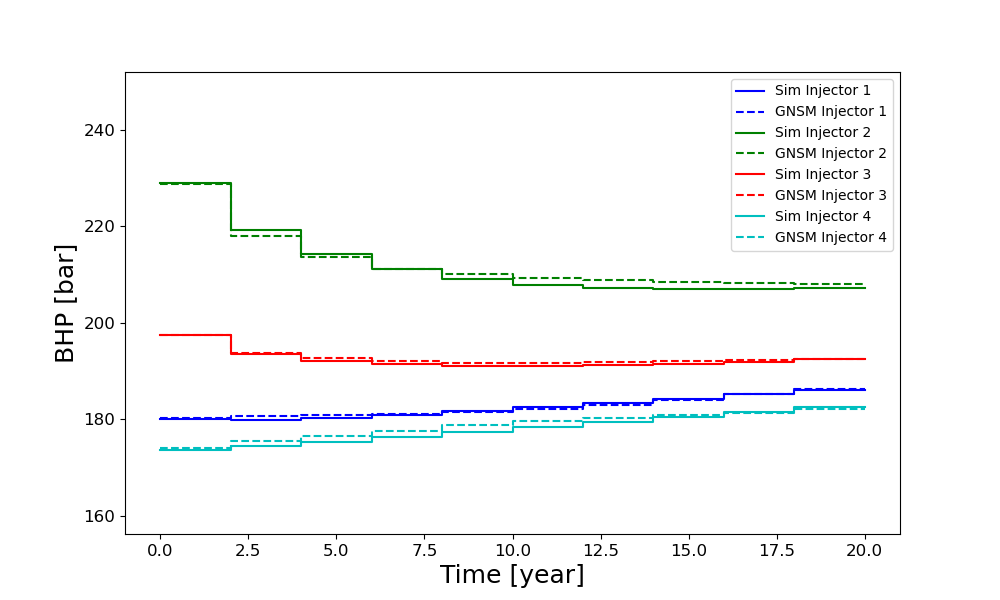}}\hfill
    \caption{BHP profiles from simulation and GNSM \& MLP for test cases~1 and~2.}
	\label{fig:bhp_test12_3d}
\end{figure}

\begin{figure}[H]
	\centering
	\subfloat[BHP for test case~3]{\includegraphics[width=0.9\textwidth]{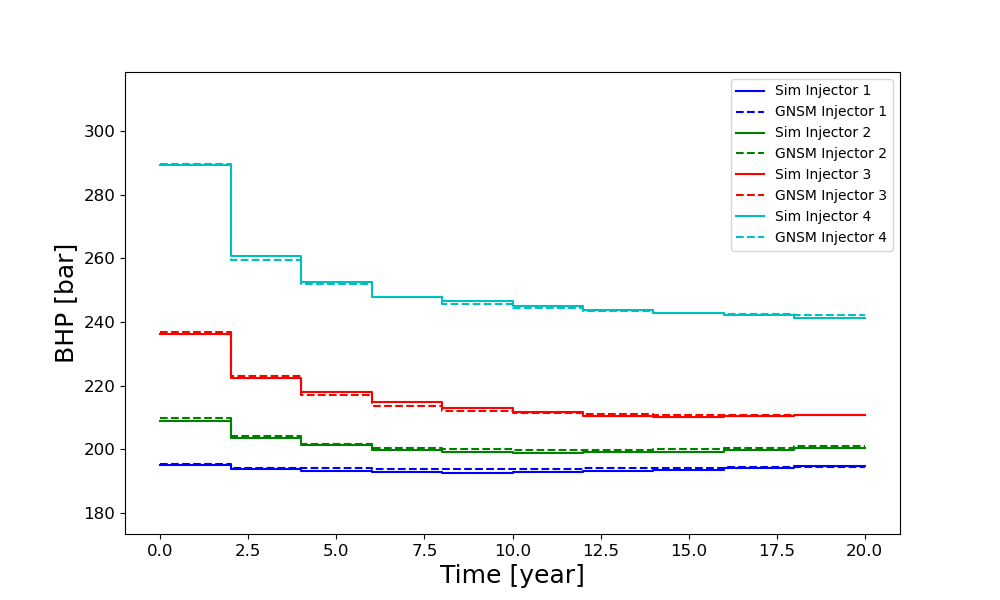}}\hfill
	\subfloat[BHP for test case~4]{\includegraphics[width=0.9\textwidth]{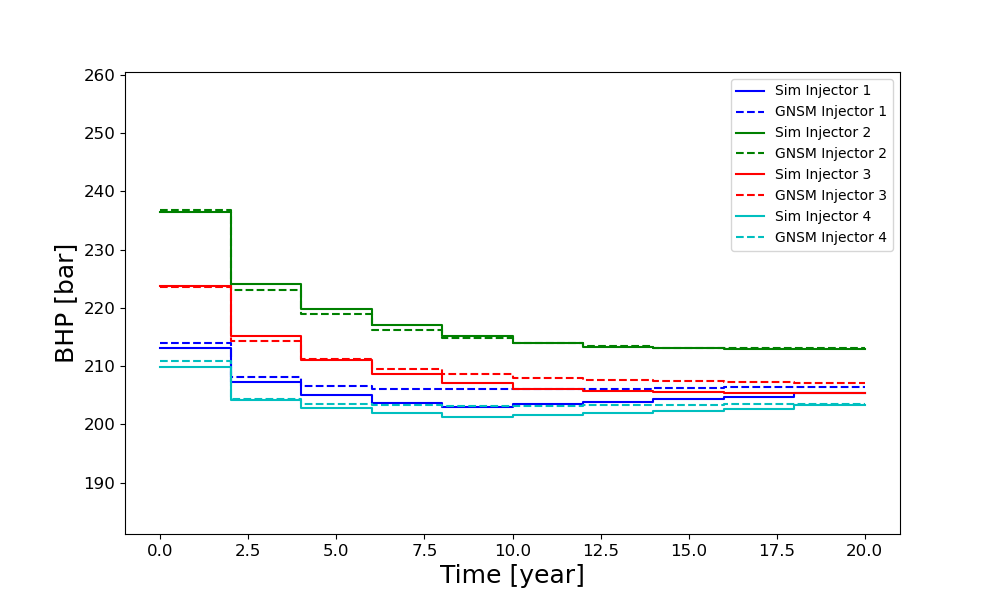}}\hfill
    \caption{BHP profiles from simulation and GNSM \& MLP for test cases~3 and~4.}
	\label{fig:bhp_test34_3d}
\end{figure}

As explained earlier, BHPs are predicted with a separate MLP that uses GNSM results and other features as inputs. BHP profiles, in time for each well, are shown in Fig.~\ref{fig:bhp_test12_3d} (test cases~1 and 2) and Fig.~\ref{fig:bhp_test34_3d} (test cases~3 and 4). In these figures, the solid and dashed lines represent the BHP of each horizontal injector from simulation and GNSM, respectively. Each color corresponds to a different well. A high degree of accuracy is achieved by the MLP for all wells in all four cases, even though the well BHPs vary from about 175~bar to about 290~bar. We also observe different trends in time for the various wells, which are in all cases captured by the MLP. BHP prediction is important for simulation and optimization, and it is evident from Figs.~\ref{fig:bhp_test12_3d} and~\ref{fig:bhp_test34_3d} that our GNSM is able to provide accurate estimates of this essential quantity.

\begin{figure}[H]
	\centering
	\subfloat[Example~1]{\includegraphics[width=0.5\textwidth]{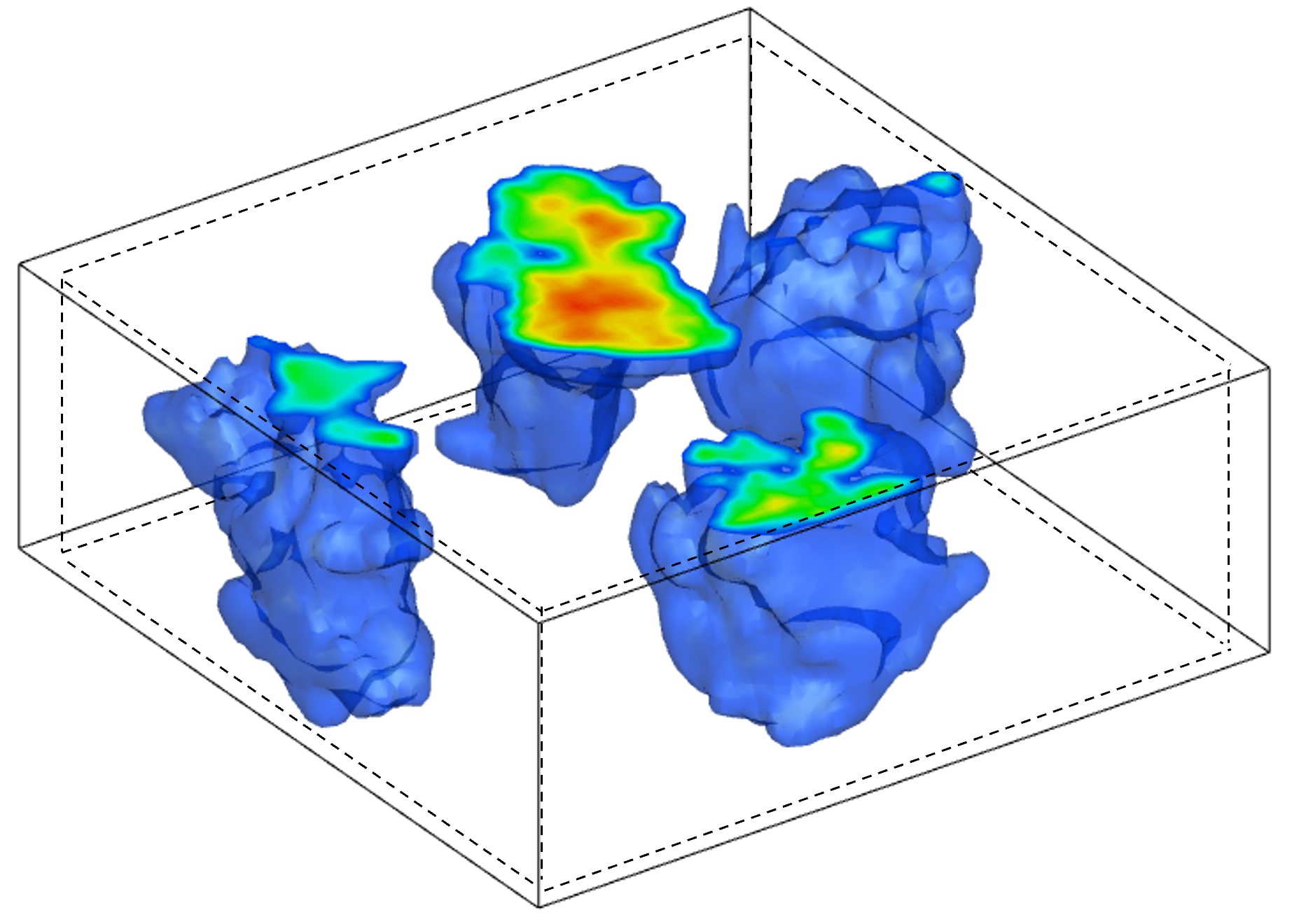}}\hfill
	\subfloat[Example~2]{\includegraphics[width=0.5\textwidth]{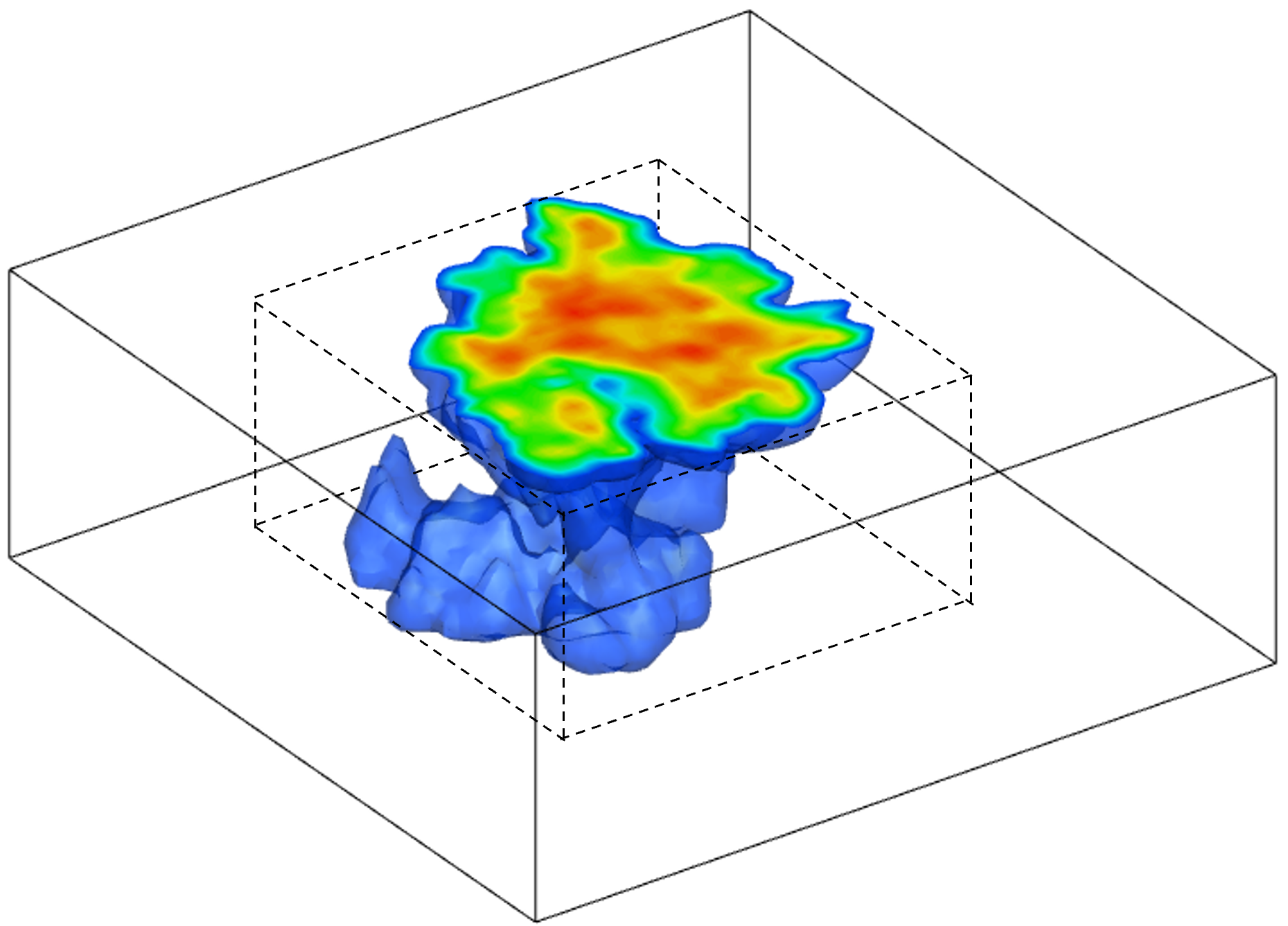}}\hfill
    \caption{Illustration of CO$_2$ footprint for two cases. Volume of the box-shaped region (dashed lines) enclosing the plumes corresponds to $V_{fp}$.}
	\label{fig:footprint_illustration}
\end{figure}

Our final evaluation involves the assessment of the accuracy of the GNSM for the calculation of the CO$_2$ footprint after 20~years of injection. This quantity is important in this work because this is the objective function that will be minimized in our optimizations. The CO$_2$ footprint for test case $i$ (or for candidate optimization solution $i$) is denoted $V_{fp}^i$. This quantity is simply the box-shaped bulk volume necessary to contain the entire CO$_2$ plume for sample $i$. Saturation values below a threshold are set to zero in this computation. Specifically, we do not consider cells with $S_g \leq 0.05$ as contributing to the footprint. To compute $V_{fp}^i$, we find the minimum 3D box-shaped volume that contains the plume in all layers. The box cross-section in the $x$-$y$ plane is aligned with the coordinate axes. The volume is extended vertically to contain the total thickness of the storage aquifer. The computation of $V_{fp}^i$ is illustrated in Fig.~\ref{fig:footprint_illustration} for two example cases. The box-shaped volumes are indicated by the dashed lines.

\begin{figure*}[!htb]
\centering
\includegraphics[width = 0.7\textwidth]{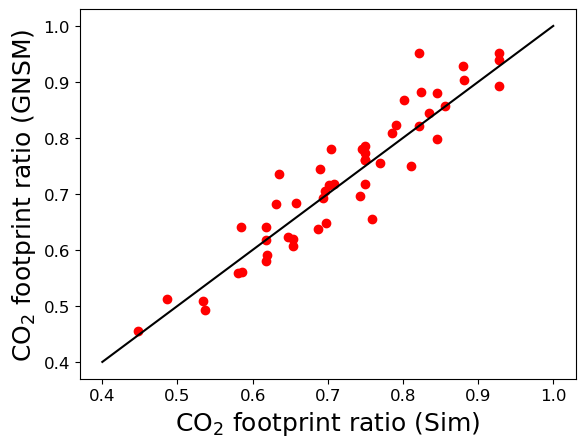}
\caption{Cross-plot of CO$_2$ footprint ratio computed from GNSM and simulation results. The 45-degree line represents perfect agreement.} \label{fig:footprint_3d}
\end{figure*}

A cross plot of the CO$_2$ footprints from GNSM and simulation, for the 50 test cases, is shown in Fig.~\ref{fig:footprint_3d}. The footprint volumes are normalized by the total bulk volume of the storage aquifer $V_{tot}$, i.e., for model $i$ we display the footprint ratio, given by $R^i_{fp} = V^i_{fp}/V_{tot}$. The 45-degree line represents perfect agreement. The general trend is clearly captured by GNSM, though there is some scatter in the results. These results, along with the BHP predictions shown in Figs.~\ref{fig:bhp_test12_3d} and~\ref{fig:bhp_test34_3d}, suggest that the GNSM will be suitable for the constrained optimization problem considered in the next section.

\section{GNSM for well placement optimization}
\label{sec:result_opt_3d}

In this section, we first describe the optimization problem and constraints considered in this study. Results for both simulation-based and GNSM-based optimization are then presented.


\subsection{Optimization setup}
\label{sec:opt_setup_3d}

The optimization problem can be stated as
\begin{align}
\begin{cases}
\underset{\mathbf{u}\in \mathbb{U}}{\min} \;\; \mathit{J}(\mathbf{u}),  \\
\mathbf{c}(\mathbf{u}) \leq \mathbf{0}.  \label{eqn:wpo_def}
\end{cases}
\end{align}
Here $J$ is the objective function we seek to minimize, taken here to be the CO$_2$ footprint ratio (i.e., $J=R_{fp} = V_{fp}/V_{tot}$). The optimization variables, $\mathbf{u}\in \mathbb{U} \subset{\mathbb{R}^{6n_w}}$, include the $x$, $y$ and $z$ locations of the heel and toe for each of the $n_w$ wells. Because we specify $n_w=4$ in this work, there are a total of 24 optimization variables. The space $\mathbb{U}$ defines the feasible region for $\mathbf{u}$. The nonlinear constraints for the optimization, which include geometric constraints on the well configuration (given in Table~\ref{tab:opt_constraint_3d}), BHP constraints, and a retention constraint to ensure the CO$_2$ stays in the storage aquifer, are represented by the vector $\mathbf{c}$. When $\mathbf{c}(\mathbf{u}) \leq \mathbf{0}$, all these constraints are satisfied and the solution is feasible. Note that we do not introduce any well control optimization variables, meaning all wells inject the same amount of CO$_2$ (0.5~Mt/year).

The optimizations are accomplished using differential evolution (DE), which is a population-based global stochastic search procedure. Please see~\citep{zou2022effective, zou2023integrated} for a description of DE in the current setting. We specify the population size to be the same as the number of optimization variables (24). Optimization runs are terminated when the relative change in the objective function value is less than 1\% over 20 iterations, or if a total of 50 iterations is performed. The maximum number of function evaluations with these specifications is thus $24 \times 50=1200$. 

The geometric constraints on the well configuration are provided in Table~\ref{tab:opt_constraint_3d}. Constraint~5 ensures that all wells are horizontal. Note that this could also be accomplished by formulating the problem with five optimization variables per well instead of six. We use the repair procedure introduced by Volkov and Bellout~\citep{volkov2018gradient}, and previously applied in~\citep{zou2023integrated}, to improve infeasible well configurations.



\begin{table}[htb!]
\centering
\caption{Well geometric constraints applied in optimization}
\label{tab:opt_constraint_3d}
\begin{tabular}{|c|c|c|}
\hline
{No.} &{Constraint type} & {Number} \\ \hline
{1.} &Maximum well length         &  1200~m \\
    {2.} &Minimum well length         &  480~m \\
{3.} &Minimum interwell distance  &  720~m \\
{4.} &Minimum well-to-boundary distance   & 424~m \\ 
{5.} &Maximum heel-to-toe difference in depth   & 0~m \\ \hline
\end{tabular}
\end{table}

Two additional nonlinear constraints, involving a maximum BHP and retention of CO$_2$ in the storage aquifer, enter the formulation. The BHP constraint, $C_{bhp}$, is written as
\begin{equation}\label{eq:bhp_constraint_3d}
    C_{bhp} = \frac{\max(0,p^{bhp}_{max} - p^{bhp}_{allow})}{p^{bhp}_{allow}},
\end{equation}
where $p^{bhp}_{max}$ is the maximum BHP value, for any well at any time step, observed for the current well configuration, and $p^{bhp}_{allow}$ is the specified maximum allowable BHP. Consistent with Crain et al.~\citep{crain2024integrated}, we set $p^{bhp}_{allow}=276.3$~bar. The CO$_2$ retention constraint, $C_{ret}$, is expressed as 
\begin{equation}\label{eq:retention_constraint_3d}
    C_{ret} = \frac{m^{inj}-m^{ret}}{m^{inj}},
\end{equation}
where $m^{inj}$ is the total mass of CO$_2$ injected over 20~years (which in our case is 40~Mt), and $m^{ret}$ is the mass of CO$_2$ retained in the storage aquifer. Any CO$_2$ that is not retained in the storage aquifer will appear in the outer region shown in Fig.~\ref{fig:reservoir_model_3d}b.

In our optimization framework, the BHP constraint (Eq.~\ref{eq:bhp_constraint_3d}) and the CO$_2$ retention constraint (Eq.~\ref{eq:retention_constraint_3d}) are handled independently using the filter method. This approach, which was also used in~\citep{zou2023integrated}, treats constraint violations as additional objectives to be minimized. Minimization of constraint violations is prioritized over minimization of $J$ when constraints are violated. For both $C_{bhp}$ and $C_{ret}$, we set a feasibility threshold of $10^{-5}$.

\subsection{Optimization results}
\label{sec:opt_results_3d}

\begin{figure*}[!htb]
\centering
\includegraphics[width = 0.7\textwidth]{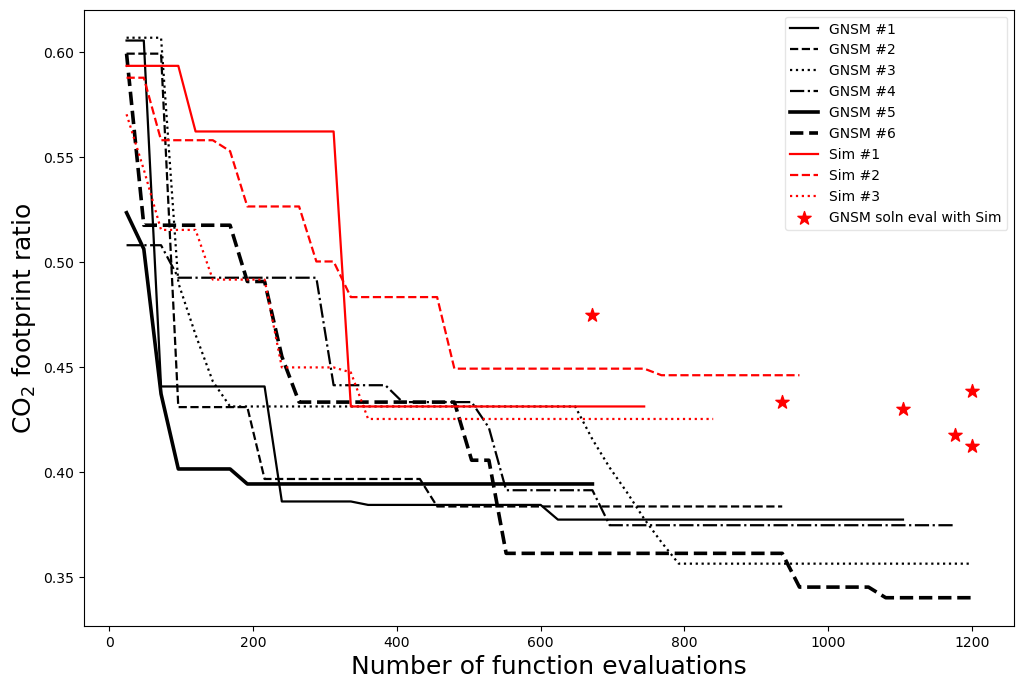}
\caption{Progress of optimizations for simulation-based and GNSM-based runs.} \label{fig:optimization_progress_runs_all_simulation_3d}
\end{figure*}

DE and other population-based optimizers are stochastic, and the optimal solution can vary significantly from run to run because they do not (in general) find the global minimum. Thus, by performing many runs with GNSM-based optimization, which can be accomplished very efficiently with full parallelization, we expect to achieve better solutions than would be found using only a few (expensive) simulation-based optimization runs. Along these lines, we perform three simulation-based and six GNSM-based optimization runs within the Stanford Unified Optimization Framework. The progress of each of these optimization runs is displayed in Fig.~\ref{fig:optimization_progress_runs_all_simulation_3d}. The simulation-based runs are shown in red, and the GNSM-based runs are in black. Results are presented in terms of the CO$_2$ footprint ratio (objective function) versus the number of function evaluations (flow simulations or GNSM evaluations) performed. The red stars in Fig.~\ref{fig:optimization_progress_runs_all_simulation_3d} indicate the CO$_2$ footprint ratio computed using Eclipse CO2STORE with the optimal well configuration found by GNSM-based optimization. The differences in footprint ratio between the stars and the black curves quantify GNSM error for the optimum solution. 

The optimal CO$_2$ footprint ratio in the simulation-based runs is 0.425, achieved in run~3. The optimal footprint ratio in a GNSM-based run is 0.412 (run~6). The Eclipse CO2STORE results for the run~6 well configuration confirm that this solution satisfies all constraints. This set of results illustrates the benefits of GNSM-based optimization.

Some of the test-set results for CO$_2$ footprint ratio in Fig.~\ref{fig:footprint_3d} are quite close to those achieved via optimization. However, in the test-set evaluations, constraint violations were not considered. For many of the low-footprint solutions in Fig.~\ref{fig:footprint_3d} there are, in fact, BHP or CO$_2$ retention constraint violations. Thus, these solutions would be rejected (as infeasible) during optimization.

\begin{figure}[H]
	\centering
	\subfloat[Simulation saturation in layer~1]{\includegraphics[width=0.4\textwidth]{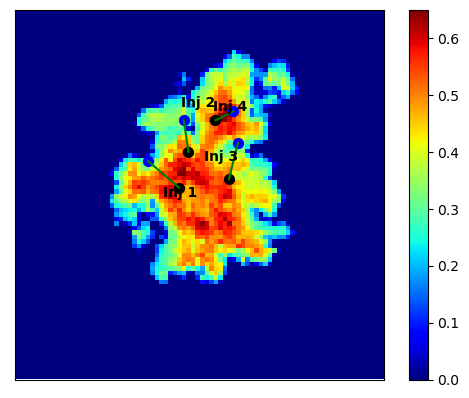}}\hfill
	\subfloat[GNSM saturation in layer~1]{\includegraphics[width=0.4\textwidth]{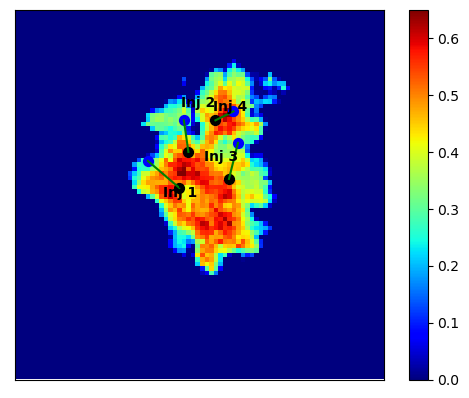}}\hfill
	\subfloat[Simulation saturation in 3D]{\includegraphics[width=0.4\textwidth]{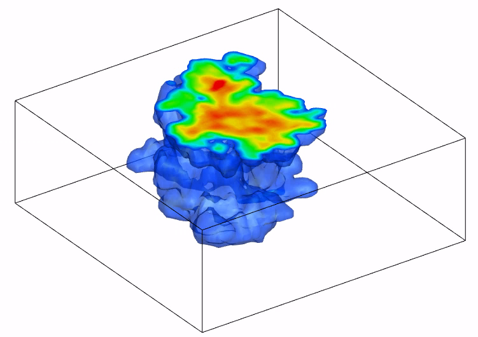}}\hfill
	\subfloat[GNSM saturation in 3D]{\includegraphics[width=0.4\textwidth]{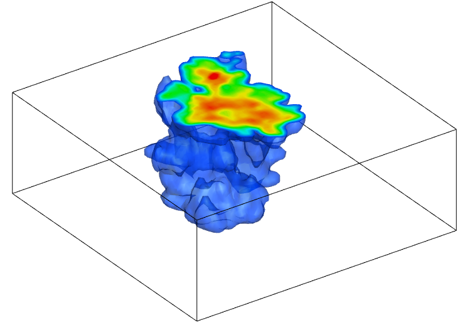}}\hfill
    \caption{Saturation fields at 20~years for simulation (left) and GNSM (right) for the optimal well configuration from the best GNSM-based optimization (run~6). Green lines in upper row display the horizontal injectors projected onto the first layer.}
	\label{fig:saturation_maps_optimal_gnn_3d}
\end{figure}

Saturation fields at 20~years corresponding to the optimal solution from GNSM-based optimization are shown in Fig.~\ref{fig:saturation_maps_optimal_gnn_3d}. The 2D and 3D fields on the left are simulation results using the well configuration found in run~6 of GNSM-based optimization, and those on the right are the SatGNN predictions. The projected wells appear in the 2D maps. Wells~1, 2, 3, and 4 are located in layers~9, 5, 7, and 1, respectively. It is apparent that the optimal plume is very compact, as would be expected since the optimization seeks to minimize the footprint. We reiterate that all constraints (geometric and dynamic) are satisfied by this solution. There are some differences between the GNSM and simulation solutions, with the simulated plume slightly larger than the GNSM prediction (consistent with Fig.~\ref{fig:optimization_progress_runs_all_simulation_3d}), though the two solutions correspond closely in an overall sense.

We now discuss optimization timings. As noted earlier, a single Eclipse CO2STORE simulation requires about 8~minutes, while a GNSM evaluation takes about 4~seconds per prediction. This GNSM time includes 1.9~seconds for overhead tasks such as importing libraries, loading models, and constructing graph data (this overhead could potentially be reduced with a better implementation), and 2.1~seconds for the GNSM prediction itself. 
In terms of runtime computations, the total sequential time for one optimization run requiring 1200 flow simulations would be 160~hours, while one such run using GNSM would require about 1.3~hours. These timings would of course both be reduced considerably via parallelization. We note that there are some complications associated with full parallelization of GNSM-based optimization within our Unified Optimization Framework. This is due to complexities that arise when CPU and GPU modules are combined in HPC settings, including memory sharing issues in GPUs. As a result, observed speedups may be less than a factor of 120 with our current implementation.

Speedup here is larger than that observed for the 2D oil-water systems considered in~\citep{tang2024graph}. This is because our GNSM scales well with increasing model size, because we predict for fewer time steps in the present case, and because a single-phase pressure solution, which was used as a GNSM feature in~\citep{tang2024graph}, is not used here.

\section{Concluding remarks}
\label{sec:summary}

In this study, we developed a graph network surrogate model (GNSM) able to predict 3D pressure and saturation fields in a storage aquifer, along with BHP for each well, for new configurations involving four horizontal CO$_2$ injection wells. The model provides these results at a specified set of time steps. The GNSM involves separate networks, with the same architecture but characterized by different hyperparameters, for the prediction of pressure and saturation. BHPs are predicted in a separate step using an MLP, which accepts GNSM results as input. Results for new test cases demonstrated the capabilities of the surrogate model. Median errors for pressure and saturation predictions were about 4\% and 6\%, respectively. BHP predictions were also shown to be very accurate. The GNSM performed well for cases with very different plume structures, as would be encountered during an optimization run.

The GNSM was applied for well placement optimization, where the objective was to find the well configuration that minimizes the CO$_2$ footprint. Both geometric and dynamic constraints were imposed. GNSM-based optimization provided results comparable to those from simulation-based optimization, but the GNSM procedure involves much faster function evaluations. Specifically, a single flow simulation required about 8~minutes, while a GNSM function evaluation took about 4~seconds, which corresponds to a speedup of $120\times$. This level of speedup was not achieved in the full optimization, however, due to parallelization issues within the Stanford Unified Optimization Framework.

In future work, we can consider extending the GNSM to handle different numbers of injectors, variable well controls (e.g., time-varying injection rates) and deviated wells. This added flexibility should lead to better optimized solutions. Other objective functions, such as minimizing mobile CO$_2$, or biobjective optimization with mobile CO$_2$ and CO$_2$ footprint as objectives, should also be addressed. The ability of the GNSM to provide predictions over multiple geomodels should be evaluated and enhanced as required, as this could facilitate its use for robust optimization. Finally, the GNSM should be tested, and extended as necessary, to treat larger and more realistic carbon storage models.

\section*{CRediT authorship contribution statement}
{\bf Haoyu Tang:} Conceptualization, Methodology, Software, Formal analysis, Visualization, Writing – original draft. {\bf Louis J.~Durlofsky:} Conceptualization, Formal analysis, Resources, Writing – review and editing.

\section*{Declaration of competing interest}
The authors declare that they have no known competing financial interests or personal relationships that could have appeared to influence the work reported in this paper.

\section*{Data availability}
Please contact Haoyu Tang (hytang@stanford.edu) to request the data, models and code used in this study.

\section*{Acknowledgements}
We are grateful to GeoCquest II, a BHP-sponsored collaborative project involving the University of Melbourne (Australia), the University of Cambridge (UK), and Stanford University (USA), for funding. We thank Su Jiang for supplying the geological model used in this study, and Oleg Volkov and Amy Zou for their assistance with the Stanford Unified Optimization Framework. We also acknowledge the SDSS Center for Computation for providing HPC resources.

\end{document}